\documentclass{article}

\usepackage{arxiv}

\usepackage[utf8]{inputenc} 
\usepackage[T1]{fontenc}    
\usepackage{hyperref}       
\usepackage{url}            
\usepackage{booktabs}       
\usepackage{amsfonts}       
\usepackage{nicefrac}       
\usepackage{microtype}      
\usepackage{lipsum}

\usepackage{graphicx}
\usepackage{adjustbox}
\usepackage{caption}
\usepackage{subcaption}
\usepackage{algorithm}
\usepackage[noend]{algpseudocode}
\usepackage{siunitx} 
\usepackage{gensymb}
\usepackage{rotating}
\usepackage{multirow} 

\usepackage{amssymb,amsmath}

\usepackage{lineno}

\title{Towards smart sustainable cities: Addressing semantic heterogeneity in Building Management Systems using discriminative models}

\author{
  Chidubem Iddianozie\thanks{corresponding author} \\
  School of Computer Science\\
 University College Dublin\\
  \texttt{chidubem.iddianozie@ucdconnect.ie} \\
   \And
Paulito Palmes \\
  IBM Research\\
 Ireland \\
  \texttt{paulito.palmes@ie.ibm.com} \\
}

\begin{document}
\maketitle

\begin{abstract}
Building Management Systems (BMS) are crucial in the drive towards smart sustainable cities. This is due to the fact that they have been effective in significantly reducing the energy consumption of buildings. A typical BMS is composed of \textit{smart} devices that communicate with one another in order to achieve their purpose. 
However, the \textit{heterogeneity} of these devices and their associated meta-data impede  the deployment of solutions that depend on the interactions among these devices. Nonetheless, automatically inferring the semantics of these devices using data-driven methods provides an ideal solution to the problems brought about by this heterogeneity. In this paper, we undertake a multi-dimensional study to address the problem of inferring the semantics of IoT devices using machine learning models. 
Using two datasets with over 67 million data points collected from IoT devices, we developed discriminative models that produced competitive results. Particularly, our study highlights the potential of Image Encoded Time Series (IETS) as a robust alternative to statistical feature-based inference methods. Leveraging just a fraction of the data required by feature-based methods, our evaluations show that this encoding competes with and even outperforms traditional methods in many cases.
\end{abstract}

\keywords{AI for Buildings \and Smart sustainable cities \and IoT devices \and Building Management Systems \and Energy Management \and Time Series Classification}

\section{Introduction}
\label{introduction}

Over 50\% of the world's population live in cities~\cite{UN2018}. This trend which is expected to be on the increase for the foreseeable future will inadvertently create challenges for the infrastructure, economy and environment of cities~\cite{UN2018,akande2019lisbon}. In light of this, it becomes evident that cities are a strategic sector in the fight against global issues such as climate change. Naturally, there has been an intensified global move to ensure that urban places are more \textit{sustainable} in order to alleviate these challenges~\cite{akande2019lisbon,leal2018reinvigorating,lu2015policy,campisi2018evaluation}. An effective approach towards achieving sustainability has been by making cities \textit{smart} through the use of ICT~\cite{silva2018towards,chen2014vision}. In order to eliminate the common obfuscation between \textit{smartness} and \textit{sustainability} in the context of cities, we adopt the notion of \textit{smart sustainable} cities~\cite{ahvenniemi2017differences}. According to the United Nations Economic and Social Council, a smart sustainable city is defined as an \textit{innovative city that uses information and communication technologies (ICTs) and other means to improve quality of life, efficiency of urban operation and services, and competitiveness, while ensuring that it meets the needs of present and future generations with respect to economic, social, environmental as well as cultural aspects}~\cite{UNECE2015}.


There have been several approaches that seek to reduce energy consumption in cities using ICT. This is an important aspect supported by studies showing estimates of up to a 16.5\% reduction of total greenhouse gas emissions from ICT solutions~\cite{kramers2014smart}. Among these solutions, the Building Management Systems (BMS) are easily one of the most significant~\cite{lilis2017towards,jung2012integrating,akkaya2015iot}. A BMS can help reduce energy consumption by almost 15\% through managing the heating, lighting, ventilation, air conditioning of a building~\cite{brambley2005advanced}. 

Typically, a BMS employs Internet of Things (IoT) composed of smart devices\footnote{We use the terms \textit{IoT} devices and \textit{smart} devices interchangeably} and a communication infrastructure~\cite{akkaya2015iot}. These smart devices could be \textit{sensors} that monitor environmental phenomena or actuators that perform certain actions~\cite{lilis2017towards}. In addition to the benefit of reducing energy demands of buildings, a BMS can be used to improve the well-being of building dwellers and promote livability~\cite{shaikh2014review}. The significance of a BMS becomes clearer when you consider the relevance of buildings in society. It has been shown that people spend up to 90\% of their lives inside buildings~\cite{effects2018yougov}. This implies a huge cost with providing services 
to make buildings habitable and comfortable such as heating, ventilation, and lighting~\cite{shaikh2014review}. 
In the US alone, buildings consume 30\% of the energy generated and make up nearly 41\% of the annual energy budget~\cite{bhattacharya2015automated,gao2015data,mehta2013sustainable}. With this in mind, it becomes obvious how this can benefit broader critical issues such as climate change, pollution, sustainable and livable cities~\cite{mehta2013sustainable,lilis2017towards,akande2019lisbon}. Consequently, the link between Building Management Systems and \textit{smart sustainable} cities becomes too strong to ignore~\cite{bhattacharya2015automated,gao2015data,lilis2017towards}.

There is a need for the smart devices in a BMS to communicate together in order to achieve a shared goal~\cite{lilis2017towards}. However, this communication, otherwise known as interoperability can be impeded by heterogeneity. Indeed, this has been identified as one of the key challenges in deploying a true smart sustainable city~\cite{silva2018towards,lilis2017towards}.

In this paper, we associate the heterogeneity of an IoT device with the data generated by that device. Typically, IoT devices are labelled using meta-data \cite{gao2015data}. An example is \textit{B1\_Temp\_Sensor} which refers to a \textit{sensor} device for \textit{temperature} in \textit{Building 1}. In this paper, we refer to this labelling as  the semantic of the device and recognise two categories of heterogeneity. 1) The first one is brought about by the different naming conventions employed by vendors and maintainers of these devices \cite{gao2015data,koh2018scrabble}. For example, a sensor IoT device can be labeled either \textit{Sensor} or \textit{Sns} although they are the same and record the same type of phenomenon. 2) The second category is introduced by subtle differences between phenomena recorded by devices. For example, you could have two devices recording temperature readings labelled as \textit{temperature} whereas in reality they are different with one taken from a living room and the other taken from a boiler room. 

In addition to these problems, the meta-data could be incorrectly labelled, outdated or not available~\cite{gao2015data}. Bearing all these in mind, we further clarify the challenges by considering a BMS composed of multiple sensing devices in two scenarios. In the first scenario, the BMS is processing IoT data with either an incorrect or outdated or no labelling at all. The problem here is to associate this data with the correct semantic group and update the labelling. In the second scenario, the BMS is processing data that is labelled similarly whereas they are semantically different (see second category of heterogeneity above). Here, the problem will be to deduce this semantic difference and update the labelling to reflect this difference. In both situations, we see that these problems can be addressed by  automatically inferring the semantics of devices using data-driven methods~\cite{shi2018survey}. In our experiments, we used datasets that incorporates these heterogeneities and challenges.

There have been some attempts to address this problem using paradigms such as multi-label classification and active learning \cite{gao2015data, bhattacharya2015automated,koh2018scrabble}. However, the literature lacks a comprehensive exposition of data-driven methods in  conjunction with the posed domainal challenges. Hence, in this paper, we carry out a quantitative study of both supervised and unsupervised learning methods for inferring the semantics of IoT devices. Particularly, we propose the use of Image Encoded Time Series (IETS) as an alternative to traditional descriptive feature-based methods for model development. In addition, we compare our proposed technique to traditional methods. Our study is based on two large datasets that contain a total of over 67 million data points with 22 semantic types. 

We formulate the problem as a multi-class classification one. In our evaluations, we observed that models trained with features generated using the proposed IETS method from a fraction (one month) of the datasets outperforms feature-based techniques in many cases. This observation    
highlights the potential of IETS as an alternative to feature-based model development strategies. This can address the problem of data availability that impedes development of inference models in IoT settings. To the best of our knowledge, this is the first attempt to provide a comparative study of machine learning methods for semantic inference of IoT devices. 
Our contributions in this paper can be summarized as follows:
\begin{itemize}
    \item We carry out a study that develops and quantitatively compares supervised and unsupervised inference methods on two large IoT datasets.
    \item We propose Image Encoded Time Series (IETS) and demonstrate that it can be a viable alternative to traditional feature-based methods for model development. 
\end{itemize}
The rest of this paper is organized as follows. Section~\ref{relatedWork} discusses related work. We describe the proposed IETS method in Section~\ref{IETS} Section~\ref{methods} describes the methods and materials used. Results are presented and discussed in Section~\ref{results} followed by our conclusions in Section~\ref{conclusions}.







\section{Related Work}
\label{relatedWork}

Several approaches have been proposed to address the problem of semantic heterogeneity in IoT devices. Project Haystack was developed to standardize the naming conventions of devices in BMS~\cite{haystack2020}. This convention is yet to be widely adopted which may be explained by the fact that it struggles to recognise the vast array of IoT devices and semantics in the wild. More so, with legacy systems, it may be cheaper or feasible to maintain existing naming conventions~\cite{lilis2017towards}. Hence, the focus of recent approaches to automate the inference of these semantics. An approach to automate the deployment of an energy management system by automatically inferring their semantic labels was proposed in~\cite{schumann2014towards}. The authors demonstrated their proposal by utilizing liguistic techniques and computing the semantic similarity values of the labels. In~\cite{koc2014comparison}, the authors sought to infer the location of sensors in a building. Particularly, they focused on evaluating the feasibility of linear correlation measures over spatial dependency measures. In \cite{gao2015data}, the authors proposed a data-driven framework to infer the semantics of building systems. Similar to our approach, they considered the meta-data of these devices as their semantics. The work in~\cite{gao2015data} is the closest in comparison to our work in this paper. 

Inference methods for IoT data is related to methods the field of signal processing such as time series classification. 
Time series classification methods can be grouped into instance-based and feature-based methods \cite{fulcher2014highly,hyndman2015large,kolozali2016effect,ye2009time}. Instance-based classification simply refers to the matching of new time series data to a previously identified class based on an encoding of the data stream. An example of this paradigm is time series shapelets \cite{ye2009time}. Here, the authors proposed the use of a segment of a time series that is representative of its class, called a shapelet. While this idea was interesting in theory and plausible on small datasets, identifying an adequately representative shapelet can easily become computationally expensive on large datasets. Furthermore, \cite{hatami2018classification} performed a comparative evaluation of methods and demonstrated that shapelets may not be suitable for classification of time-series images.

Another example of instance-based classification is described in \cite{calbimonte2012deriving}. The authors proposed a method to infer semantics of time series data by using depictive slopes of their linear approximations to determine their semantics. Feature-based classifications transform the time series to a vector of features derived from the data itself which is used to make classifications \cite{fulcher2014highly}. Our work in this paper is an example of feature-based classification.


The literature on classification using time series images is still developing. In~\cite{gonzalez2018beats}, the authors proposed an encoding technique for time series data that mitigates the effect of data drifts. Paparrizos et al.~\cite{paparrizos2019grail} introduced GRAIL - an algorithm for representation learning on time series. While both approaches are methodologically similar to ours, their approaches are only compared with instance-based methods. 
Whereas, we compare with feature-based methods.
Hatami et al.~\cite{hatami2018classification} performed classification of time series images using neural networks. They transformed the time series data to a 2$-dimensional$ array using recurrence plots \cite{eckmann1995recurrence}. In comparison with other methods for time series classification, their approach outperformed state of the art methods in more than 50\% of cases. Wang et al.~\cite{wang2015encoding} proposed two new encoding strategies for time series data: Gramian Angular Fields and Markov Transition Fields. These techniques produced very good results and exhibited competitive results when evaluated across standard datasets. Our work marries advances in time series imaging to semantic inferences of IoT. Furthermore, we comprehensively evaluate the performance of discriminative algorithms across multiple dimensions. To the best of our knowledge, this is the first body of work to carry out a study of this nature.

\section{Image Encoded Timeseries (IETS)}
\label{IETS}
The goal of IETS is to derive an alternate representation for time series data using a transformation function, where this new representation will be used as input features to a model. 
For this purpose, we considered two encoding strategies: Recurrence \cite{eckmann1995recurrence} and Gramian Angular field \cite{wang2015encoding}. These encodings have produced competitive results with state of the art \cite{hatami2018classification}.  However, in our initial experiments, we observed that the \emph{recurrence encoding} performed better. Thus, this paper will only discuss the \textit{recurrence} encoding strategy and the results derived using it. 

Recurrence is a common characteristic of many physical systems. For example, one would expect the outside temperature readings to recur at certain points in time. The recurrence plot was proposed  by Eckmann et al.~\cite{eckmann1995recurrence} as a visualization tool. The basic premise of the recurrence plot is to portray the times which recognised states in the systems recur. The first step is to extract the $m-$dimensional phase space trajectories from the original time series. For the purpose of this study, we set the dimensions to 2. The output of the recurrence function is a square matrix where each cell holds the times where states recur in the system as represented by the trajectories. The recurrence plot is formally expressed as follows:
\begin{equation}
\label{recurrencePlot}
R_{i,j} = \Theta (\epsilon - \parallel \overrightarrow{x_i} - \overrightarrow{x_j}\parallel), \text{ } \overrightarrow{x_i} \in \mathbb{R}^{m}, i,j = 1,\ldots,N
\end{equation}
where, $N$ is the number of considered states $\overrightarrow{x_i}$,$\epsilon$ is a threshold distance, $\parallel\cdot\parallel$ is a norm and $\Theta(\cdot)$ is the Heaviside function. In our experiments, we restricted the number of observations for each instance to 720 time points which corresponds to one month's worth of observations. 

\section{Materials and Methods}
\label{methods}



\subsection{Dataset}
We used two datasets: the first is an openly available dataset called the REFIT smart home dataset \cite{Firth2017}. This dataset holds readings collected from 20 UK homes between 2013 and 2015. For this dataset, we considered the following types of measurements collected from sensors: Surface Temperature, Air Temperature, Actual Temperature, Air Humidity, Relative Humidity, Brightness, Motion, Intensity and Power Consumption. The second dataset was from an industrial research building in Ireland. For this dataset, we considered 12 semantic types as follows: FCU Temp Sensor, Load Energy Meter, Air Temp Sensor, Energy Meter, FCU Temp SetPoint, FCU Heating Valve CMD, FCU Fan Speed Sensor CMD, FCU Cooling Valve CMD, FCU Room Temp Sensor, FCU Flt Sensor, Temp Sensor, Light Load Energy Meter. Both datasets contain a combined total of 67 million data points. As can be seen from the meta-data of both datasets, the REFIT dataset exhibits heterogeneity across the phenomena being measured. For the industrial dataset, the heterogeneity spans both the measurements and the devices. This reflects the types of heterogeneity considered in this paper which was discussed in Section~\ref{introduction}. See Table \ref{semanticDescription} for a summary of the datasets. 


\begin{table}
\centering
  \caption{Description of the datasets. The number of data points is computed after the time interval is resampled to the hour.}
  \label{semanticDescription}
  \resizebox{0.7\textwidth}{!}{
  \begin{tabular}{lccc}
    \toprule
    \textbf{Dataset} & \textbf{No. of data points (\textit{millions})} & \textbf{No. of files} & \textbf{No. of semantic types}\\
    \midrule
   \textbf{REFIT} &10.81 & 2217 &10  \\
\textbf{Industrial} &56.86 & 1489 & 12 \\
    \bottomrule
  \end{tabular}
  }
\end{table}


\subsubsection{Preprocessing and Feature Generation}

Both datasets contained missing values and had non-uniform  time intervals. The interval of the original time recordings was less than the minute and we observed that most of the readings were redundant. 
Hence, we resampled all data recordings to the \textit{hour} by averaging.   
We handled the missing values by linear interpolation as described in~\cite{blu2004linear}. Our problem is formulated as a multi-class classification one and we encoded the semantic labels as integers. For all the experimental learning tasks, we generated three types of features. The \textit{statistical} features, \textit{IETS} features and the cosine distances. The statistical features were descriptive statistical values which we deemed discriminative enough. They included the mean, standard deviation, variance, stationarity, kurtosis, and skewness. We encoded stationarity on an ordinal scale of $0$ to $2$ where $2$ is highest stationarity. The IETS features were generated as described in Section~\ref{IETS} using the recurrence function. We used different aggregation methods to scale down the dimensionality of the features as will be discussed in Section~\ref{learningIETS}. Finally, the cosine features were generated using the cosine distances between every instance of the random variables. 


\subsection{Learning Techniques}
In this study, the inference problem is regarded as a supervised learning one. 
Each of the methods listed below was chosen after extensive preliminary experiments. For brevity, we only present and discuss the most promising methods in this paper.


\subsubsection{Supervised Learning Algorithms}
We employed six supervised learning algorithms for our experiments. These algorithms were chosen because they had the best performance in our preliminary experiments. These algorithms are Logistic Regression, 
Random Forest (RF), Decision Trees (DT), K-Nearest Neighbours (KNN), AdaBoost, and Support Vector Machines (SVM) \cite{breiman2001random,hosmer2013applied,quinlan1986induction,cover1967nearest,hastie2009multi}. We evaluated these algorithms using F-score and accuracy. Here, accuracy is the proportion (in percent) of accurately classified data. We use both the \textit{macro-} and \textit{micro-} averaged variants of the F-score. The general formulation of the F-score is as follows: 
\begin{equation}
\label{F-scoreEquation}
\textbf{F-score} = 2 * \frac{(precision * recall)}{(precision + recall)}
\end{equation}
where,
\begin{equation}
\label{PrecisionEquation}
\textbf{Precision} = \frac{(True \text{ } Positives)}{(True \text{ } Positives + False \text{ }Positives)}
\end{equation}
\begin{equation}
\label{RecallEquation}
\textbf{Recall} = \frac{(True \text{ }Positives)}{(True\text{ } Positives + False\text{ } Negatives)}
\end{equation}

\subsubsection{Clustering}
We employed unsupervised learning using clustering techniques to explore the structure of the datasets. Three clustering algorithms were used: spectral clustering \cite{ng2002spectral}, K-medoids \cite{park2009simple}, and K-means \cite{hartigan1979algorithm}. We evaluated the clustering methods using purity and entropy as described in \cite{manning2010introduction}. We define both metrics in equations \ref{ClusterPurity} and  \ref{ClusterEntropy}: 
\begin{equation}
\label{ClusterPurity}
\textbf{Purity} (\Omega,C) = \frac{1}{N}\sum_{k} \max\limits_{j}|\omega_{k} \cap c_{j}|
\end{equation}
where $\Omega = \{\omega_{1},\ldots,\omega_{k}\}$ is the set of clusters returned and $C = \{c_{1},\ldots,c_{j}\}$ is the set of classes.

\begin{equation}
\label{ClusterEntropy}
\textbf{Entropy} = H (\Omega) = \sum_{\omega \in \Omega}H(\omega)\frac{N_{\omega}}{N}
\end{equation}
where $\Omega$ and $C$ represents the set of clusters and set of classes respectively as in equation \ref{ClusterPurity}, $N_\omega$ is the size of cluster $\omega$ and $N$ is the size of the entire dataset. $H(\omega)$ is the entropy of a single cluster and is defined as follows:
\begin{equation}
\label{singleClusterEntropy}
H (\omega) = - \sum_{c \in C}\frac{|\omega_c|}{n_\omega} \log_2 \frac{|\omega_c|}{n_\omega}
\end{equation}
Here, $n_\omega$ is the size of cluster $\omega$ and for every class $c \in C$, $|\omega_c|$ is the number of type $c$ points in cluster $\omega$.


\subsubsection{Measure of Similarity}
We computed similarities between semantic types using the cosine similarity measure. Cosine similarity is defined as follows:
\begin{equation}
\label{cosineSimilarity}
\textbf{Cosine Similarity } (X_{s},X_{y}) = \frac{X_s \cdot X_y}{\parallel X_s \parallel\parallel X_y \parallel}
\end{equation}
where $X$ is a vector representation of the features. Features here represent the statistical descriptive features generated for each instance. Subscripts $s$ and $y$ two random variables.
Based on the features extracted, we computed the cosine similarities between each instance of the random variables. 

\subsection{Algorithmic Complexity}
\label{complexity}
The computational complexity of operations described in this paper for both datasets is constant. The time to generate the features is constant since we process one time-series file at each time unit. It implies that this process can be parallelized to improve speed. Similarly, the memory requirement is $O(n)$, where $n$ is the length of the time series file. The final stage of the feature generation process is to stack them into a matrix which increases the memory and storage requirement by a factor of $K$, where $K$ is the number of time series files processed. The time and memory required to train the models is $O(n)$, with $n$ denoting the number of features in that case.    

\section{Results and Discussion}
\label{results}
In this section, we present and discuss the results of our experiments as implemented by the methods outlined in Section \ref{methods}. All experiments were carried out on a Dell PowerEdge R710 Linux server with 20.5 Gb of RAM and Intel Xeon octa-core CPU @2.40GHz. 


\subsection{Exploratory Data Analysis}
\subsubsection{Clustering}
\begin{figure*}
\captionsetup[subfigure]{font=scriptsize,labelfont=scriptsize}
    \centering 
\begin{subfigure}{0.5\textwidth}
  \includegraphics[width=\linewidth]{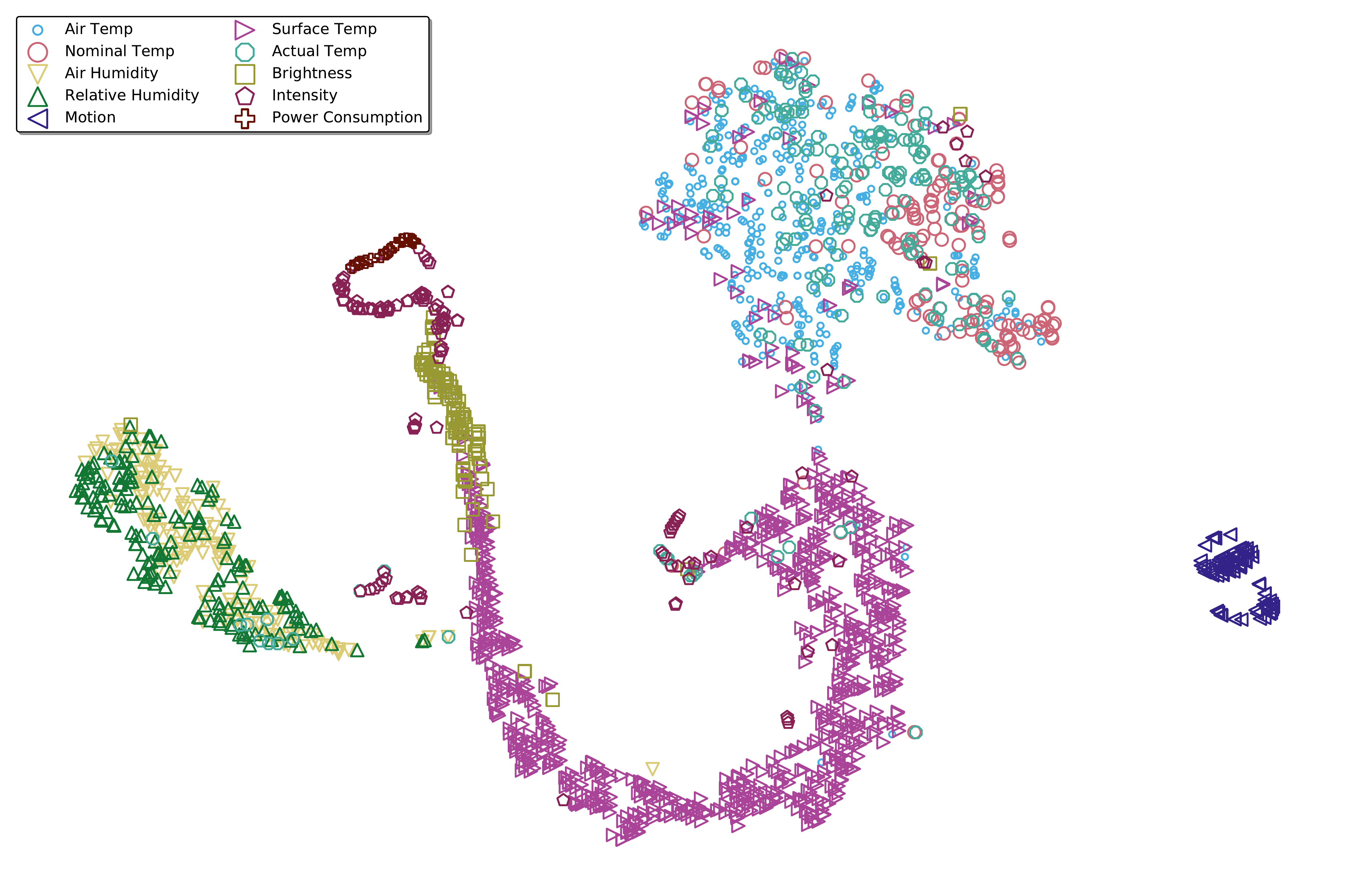}
  \caption*{Ground Truth}
  \label{fig:1}
\end{subfigure}\hfil 
\begin{subfigure}{0.5\textwidth}
  \includegraphics[width=\linewidth]{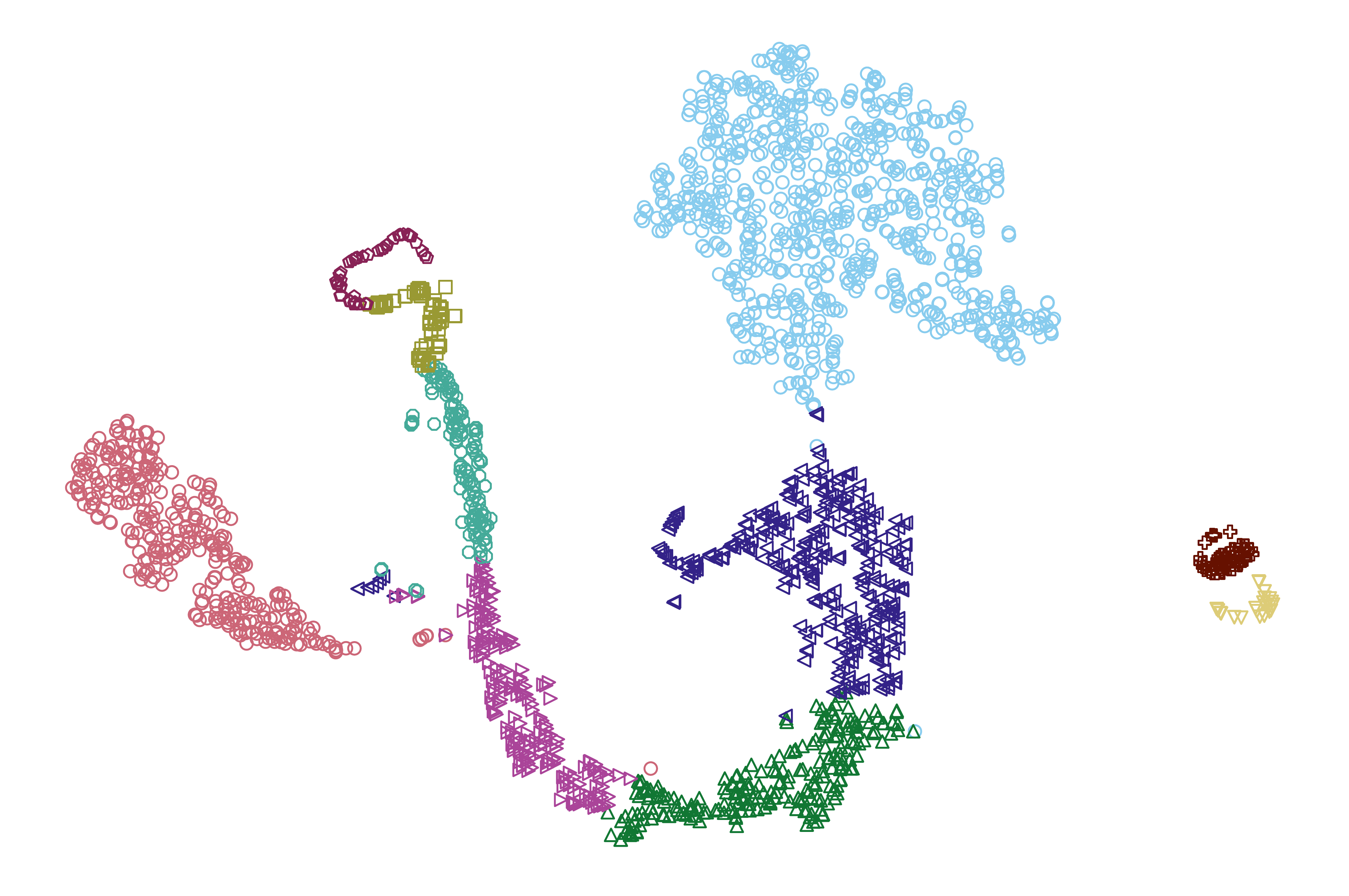}
  \caption*{Spectral Clusters}
  \label{fig:2}
\end{subfigure}\hfil 

\medskip
\begin{subfigure}{0.5\textwidth}
  \includegraphics[width=\linewidth]{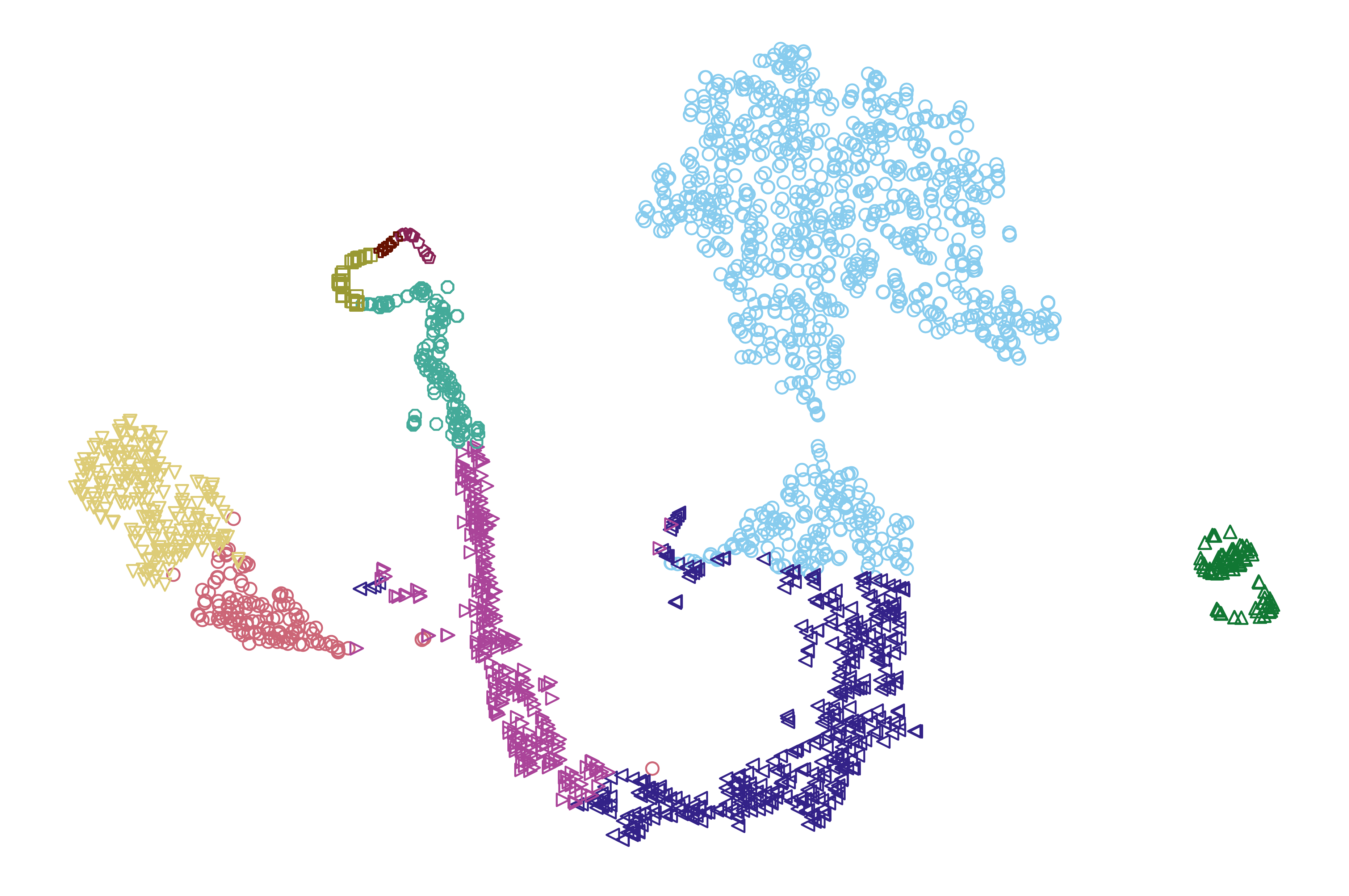}
  \caption*{K-Medoid}
  \label{fig:4}
\end{subfigure}\hfil 
\begin{subfigure}{0.5\textwidth}
  \includegraphics[width=\linewidth]{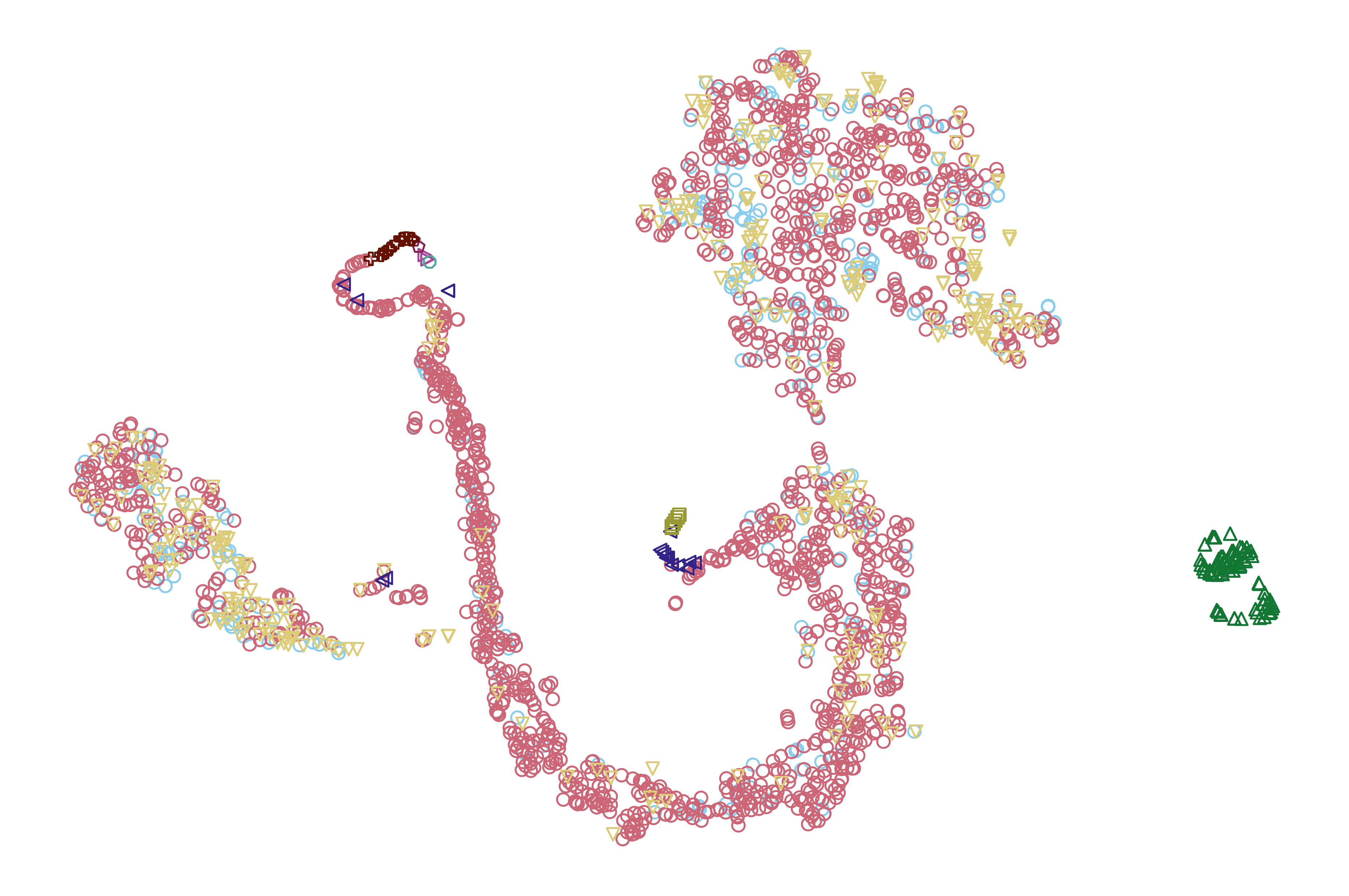}
  \caption*{K-Means}
  \label{fig:5}
\end{subfigure}\hfil 

\caption{Plots of the REFIT dataset. The first plot is that of the ground truth semantic labels. For visualization purposes, we have used the TSNE projections for the axes of all the plots. Spectral clustering performs best with a purity score of 68\% and entropy score of 1.19 compared to other clustering algorithms. See Table~\ref{clusteringAlgoPerformance} for the evaluations of the algorithms.}
\label{REFIT_plots}
\end{figure*}

\begin{figure*}
\captionsetup[subfigure]{font=scriptsize,labelfont=scriptsize}
    \centering 
\begin{subfigure}{0.5\textwidth}
  \includegraphics[width=\linewidth]{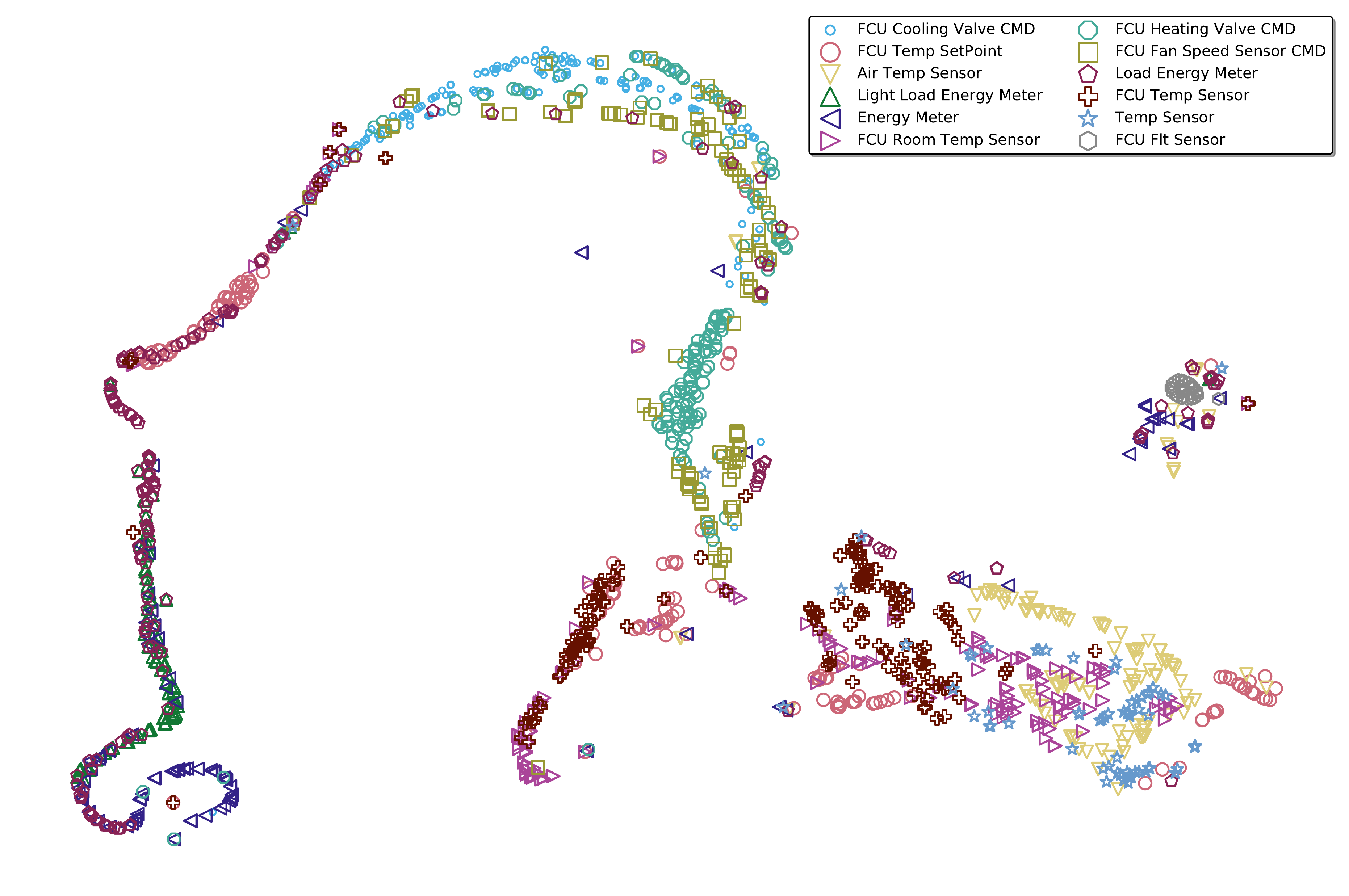}
  \caption*{Ground Truth}
  \label{fig:1}
\end{subfigure}\hfil 
\begin{subfigure}{0.5\textwidth}
  \includegraphics[width=\linewidth]{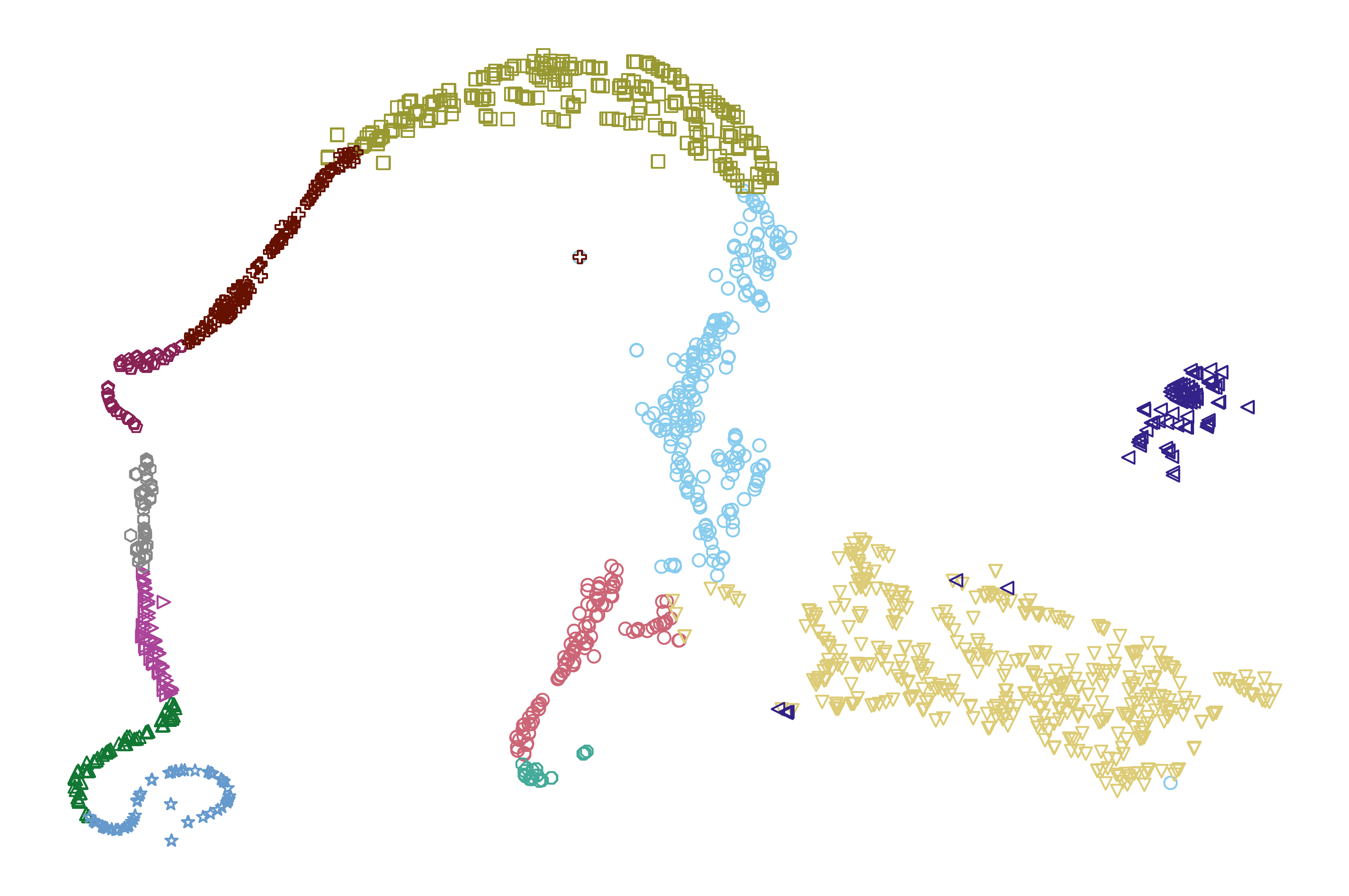}
  \caption*{Spectral Clusters}
  \label{fig:2}
\end{subfigure}\hfil 

\medskip
\begin{subfigure}{0.5\textwidth}
  \includegraphics[width=\linewidth]{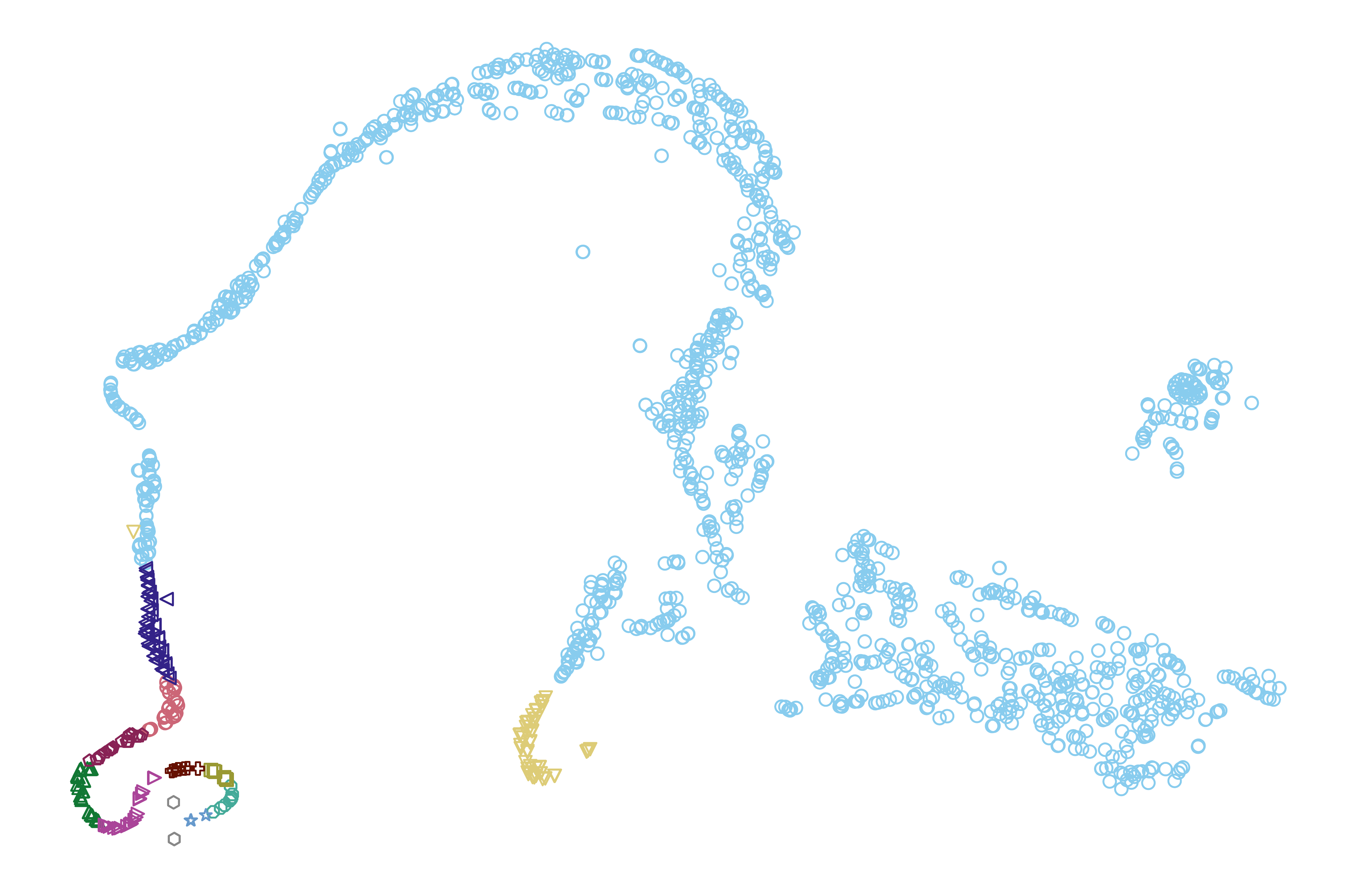}
  \caption*{K-Medoid}
  \label{fig:4}
\end{subfigure}\hfil 
\begin{subfigure}{0.5\textwidth}
  \includegraphics[width=\linewidth]{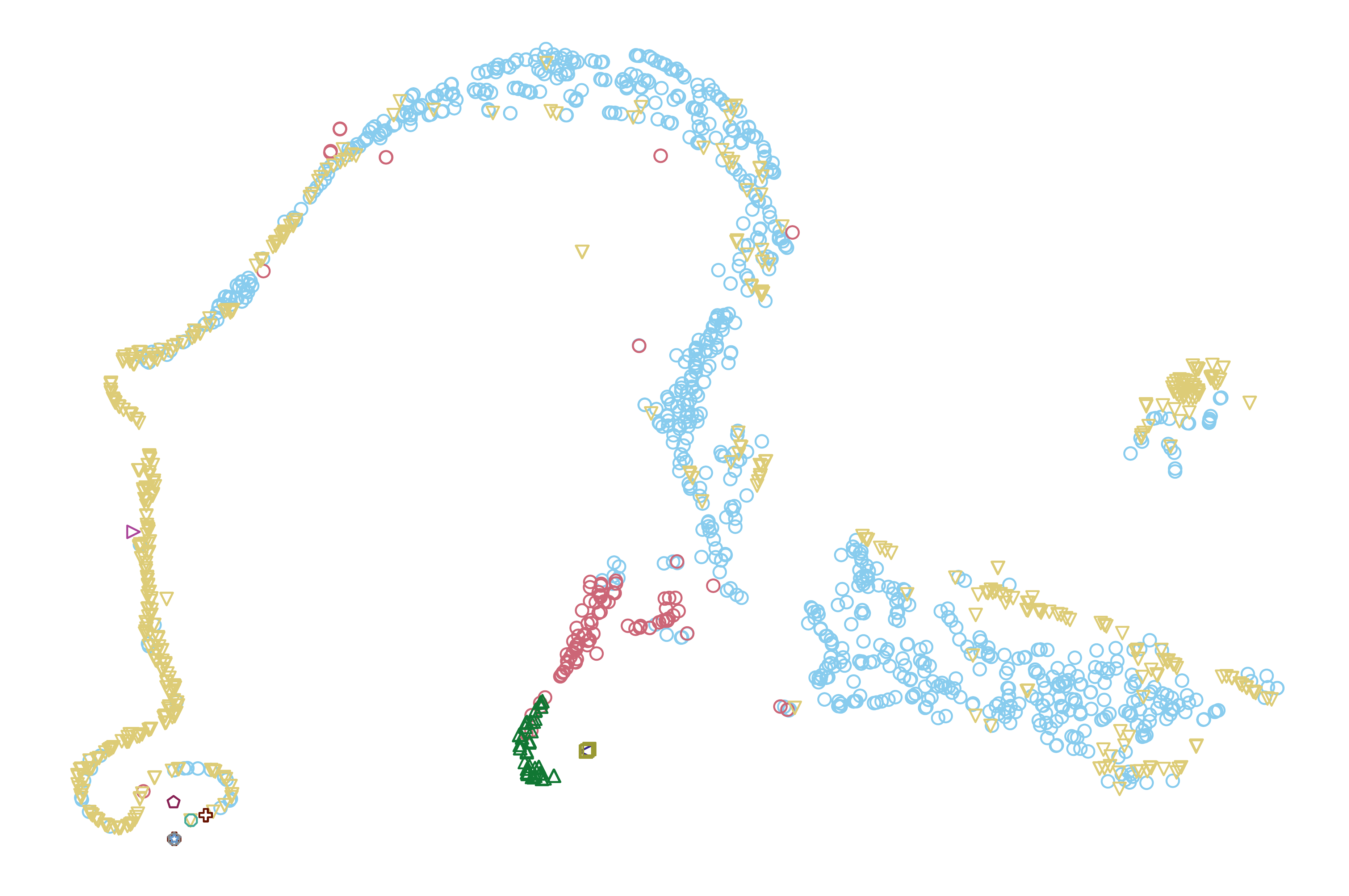}
  \caption*{K-Means}
  \label{fig:5}
\end{subfigure}\hfil 

\caption{Plots of the Industrial dataset. The first plot is that of the ground truth semantic labels. For visualization purposes, we have used the TSNE projections for the axes of all the plots. Spectral clustering performs best with a purity score of 29.3\% and entropy score of 2.0 compared to other clustering algorithms. See Table~\ref{clusteringAlgoPerformance} for the evaluations of the algorithms.}
\label{DUB_plots}
\end{figure*}

We performed exploratory analysis of the datasets using clustering methods. 
We considered three algorithms: Spectral clustering, K-medoids and K-means. The features used here are the descriptive statistical features. We evaluated the performance of these algorithms using two metrics: purity and entropy as defined in equations \ref{ClusterPurity} and \ref{ClusterEntropy}. A high percentage of purity and a low entropy score is better. We present the results of these algorithms in Figures.~\ref{REFIT_plots} and \ref{DUB_plots}. The colours in these figures represent the semantic labels plotted against the TSNE vectors on both axes \cite{maaten2008visualizing}. We can see a clear visual community-like structure to the distribution of the semantic labels using the statistical descriptive features. This would suggest that supervised methods can be successful in inferring the semantics of the IoT devices. 

Furthermore, evaluating the performance of the algorithms, as shown in Table~\ref{clusteringAlgoPerformance}, indicates that spectral clustering performs best with scores of 68\% (purity), 1.19 (entropy), 29.3\% (purity), and 2.0 (entropy) for the refit and industrial dataset respectively. Interestingly, the lower purity score recorded in the industrial dataset could be explained by the 
fact that some semantic labels in this dataset are very similar but measured by different devices. This relates back to the problems highlighted in Section~\ref{introduction}. 
\begin{table}[h!]
\centering
  \caption{Performance of the clustering algorithms. Best scores are highlighted in bold. It is seen that the spectral clustering has the best scores across both datasets.}
  \label{clusteringAlgoPerformance}
  \resizebox{0.6\textwidth}{!}{
  \begin{tabular}{lcccc}
    \toprule
    \multicolumn{1}{c}{\multirow{2}{*}{}}&
    \multicolumn{2}{c}{\multirow{1}{*}{\textbf{REFIT}}}&
     \multicolumn{2}{c}{\multirow{1}{*}{\textbf{Industrial dataset}}}\\\cmidrule(lr{1em}){2-3}\cmidrule(lr{1em}){4-5}
    
    \textbf{Algorithm} & \textbf{purity (\%)} & \textbf{entropy} &\textbf{purity (\%)} &\textbf{entropy}\\
    \midrule
   \textbf{Spectral Clustering} &\textbf{68.5}&\textbf{1.11} & \textbf{29.3} & \textbf{2.0}  \\
\textbf{K-medoids} &61.3 & 1.3 & 13.9 & 3.0 \\
\textbf{K-means} &43.3 & 2.3 &19.2 &2.8 \\
    \bottomrule
  \end{tabular}
  }
\end{table}

\subsubsection{Cosine Distances}
\label{polarPlotDistances}
\begin{figure*}[h!]
\captionsetup[subfigure]{font=scriptsize,labelfont=scriptsize}
    \centering 
\begin{subfigure}{0.5\textwidth}
  \includegraphics[width=0.9\linewidth]{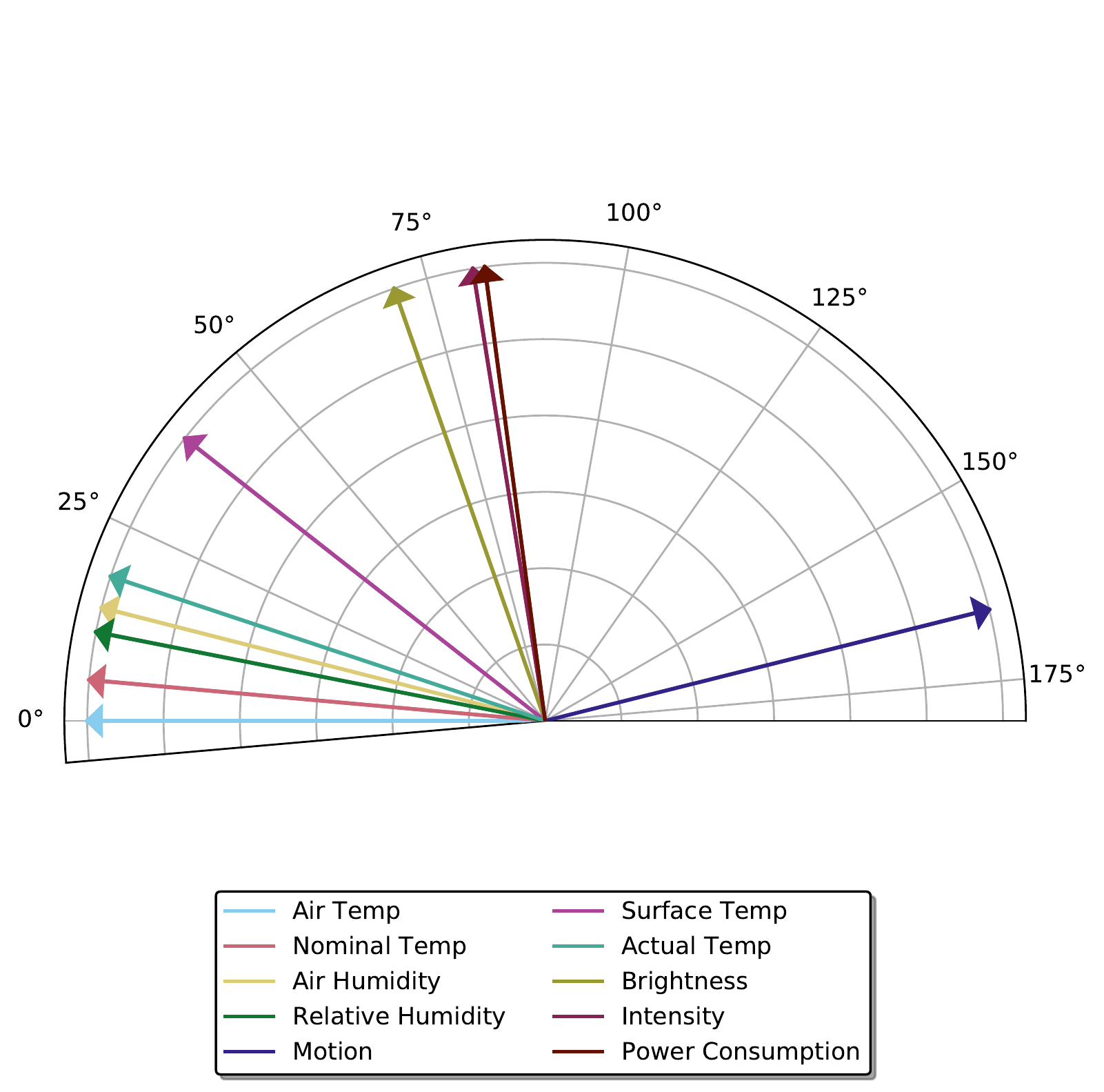}
  \caption*{REFIT}
  \label{fig:1}
\end{subfigure}\hfil 
\begin{subfigure}{0.5\textwidth}
  \includegraphics[width=0.9\linewidth]{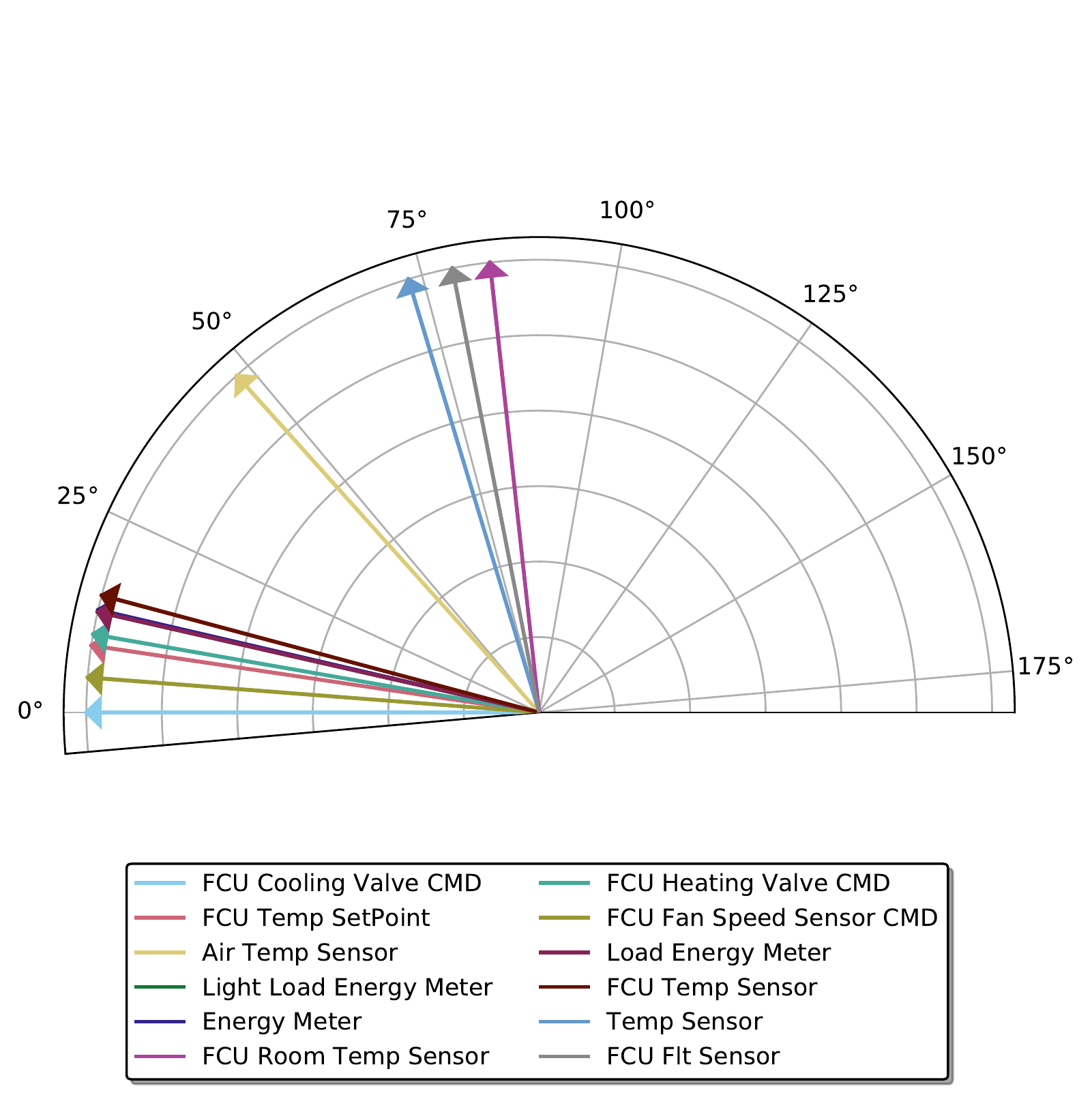}
  \caption*{Industrial Dataset}
\end{subfigure}\hfil 
\caption{Polar representation of the cosine similarity values of the semantic types present in the datasets. Each arrow represents a semantic label. The labels at 0\degree are reference points. The angular distance between any two labels denotes their similarity.}
\label{cosineSimilarityPlots}
\end{figure*}

We further explored the dataset by measuring the similarity between the semantic labels in both datasets. We computed the cosine similarity using the average of 20 random samples of each semantic label. We present the results of this analysis in Fig.~\ref{cosineSimilarityPlots}. 

In the polar plot for the REFIT dataset as shown in Fig.~\ref{cosineSimilarityPlots}, we see that the Air Temp (0\degree) is closest in similarity to Nominal Temp, then Relative Humidity, Air Humidity, Actual Temp, Surface Temp, Brightness, Intensity, Power Consumption and Motion in that order. Also, we observe that the angular distance between semantic types is comparable to the semantic similarity between the types. For example, Power Consumption and Intensity are so close together (almost completely overlapping), while there is a significant distance between motion and its closest semantic neighbour (Intensity). Thus, these observations strongly suggest that the \emph{Cosine} similarity captures the semantic similarity between the labels. 

Similarly, we consider the plot for the Industrial dataset. Here, we see that the closest label to  FCU Cooling Valve CMD (0\degree) is FCU Fan Speed Sensor CMD, then FCU Temp Set Point, FCU Heating Valve CMD, Load Energy Meter, Light Load Energy Meter, Energy Meter, FCU temp sensor, Air Temp Sensor, Temp sensor, FCU Flt sensor and FCU Room Temp Sensor in that order. 

\subsection{Supervised Learning with Descriptive Statistical Features}
\label{supervisedLearning}
\begin{figure*}
\captionsetup[subfigure]{font=scriptsize,labelfont=scriptsize}
    \centering 
\begin{subfigure}{\textwidth}
  \includegraphics[width=\linewidth]{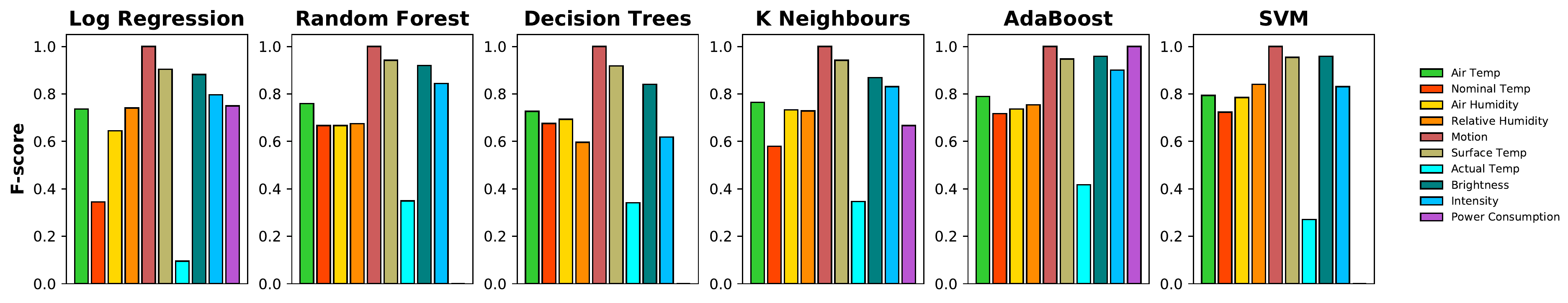}
  \caption*{REFIT}
  \label{fig:1}
\end{subfigure}\hfil 

\medskip
\begin{subfigure}{\textwidth}
  \includegraphics[width=\linewidth]{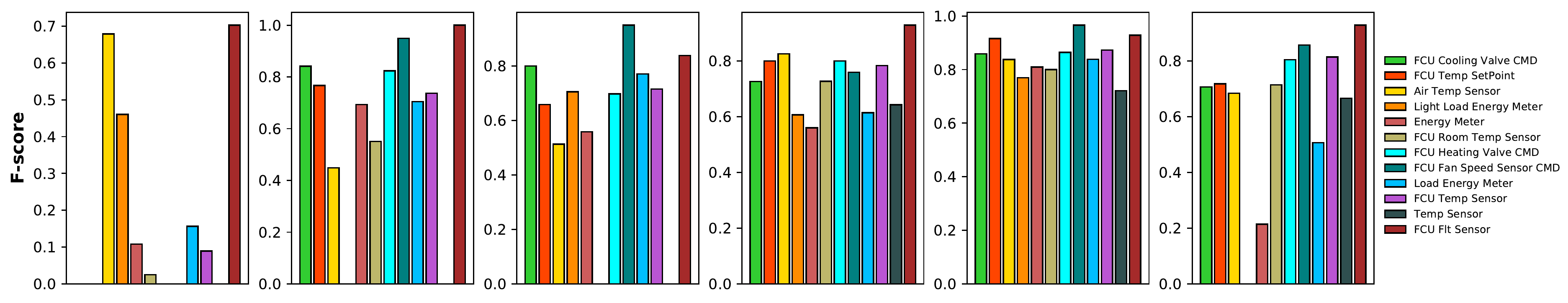}
  \caption*{Industrial Dataset}
  \label{fig:4}
\end{subfigure}\hfil 
\caption{Plots showing the F-score of the algorithms using the descriptive features extracted from the time series data. Six classical ML algorithms are used for both datasets. The sensor labels are defined by the legends to the right of each plot. It is seen that AdaBoost and KNN are the top performers for both datasets as they are able to infer all the semantics. Also, random forests performs as good for the industrial building dataset.}
\label{baseline_plots}
\end{figure*}

These experiments were carried out using the algorithms outlined in Section \ref{methods}. The results presented here are the averages taken over five runs of each model on the test set. At each run, the test is shuffled. Prior to each experiment, the class sizes for each dataset was balanced by downsampling to the smallest occurring class set. For both datasets, we split the training set and test set equally. We performed five-fold cross-validation on the training set and the best hyper-parameters were selected using a iterative search technique. The labels remain as described in Section \ref{methods}.
The features used here are the statistical descriptive features extracted from the time series data as described in Section \ref{methods}. We evaluated the models using the F-score and present the results in Fig.~\ref{baseline_plots}. The \textit{y-axis} holds the F-score values for each semantic labels. The labels for each dataset are explained using the legends to the right of each plot. 
In Fig.~\ref{baseline_plots}, we can see that for both datasets, Adaboost performs the best.
These results are very impressive, as the model achieves an F-score of 70\% in many cases. Particularly, when we consider how how semantically close/clustered the labels in the industrial dataset are as shown in Fig.~\ref{cosineSimilarityPlots}.


\subsection{Supervised Learning using Cosine Similarity Metric}
For the next investigation with supervised learning, we use the \emph{Cosine} distance as determined in Eqn.~\ref{cosineSimilarity}. Firstly, to validate their potential as described in Section~\ref{polarPlotDistances}. We observed that the cosine distances could adequately discriminate the labels in both datasets. Hence, we proceeded to use these distances as features for the supervised learning algorithms. We generated an $m \times n$ feature matrix, where $m$ is the number of time series files and $n$ is the number of semantic  labels. Then, for each semantic label, we held a \textit{signature} vector which is an average of twenty random time series files for that type. With these \textit{signature} vectors, we went ahead to compute a cosine similarity value between point and a signature vector. Thus, a matrix  cell for a particular time series file will hold the similarity value between that file and one of the \textit{signature} vectors. The models were trained using the same methodology described in Section~\ref{supervisedLearning}. We present the results in Table~\ref{methodComparisonBalanedDatasetWithStdev}.

\subsection{Supervised Learning with Image Encoded Time Series (IETS)}
\label{learningIETS}
\begin{figure*}
\captionsetup[subfigure]{font=scriptsize,labelfont=scriptsize}
    \centering 
\begin{subfigure}{0.2\textwidth}
  \includegraphics[width=\linewidth]{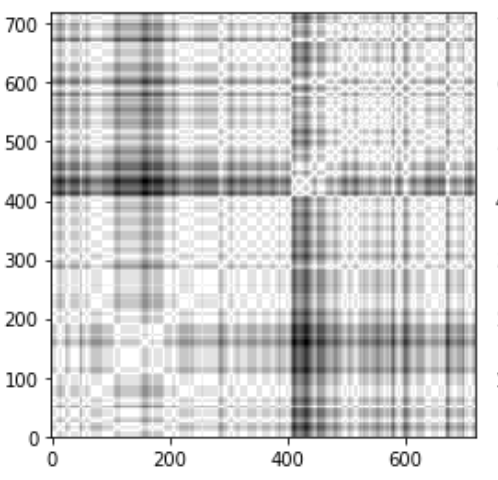}
  \caption*{Surface Temperature}
  \label{fig:1}
\end{subfigure}\hfil 
\begin{subfigure}{0.2\textwidth}
  \includegraphics[width=\linewidth]{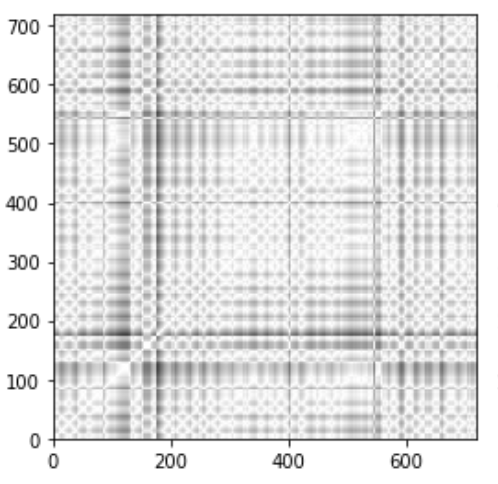}
  \caption*{Air Temperature}
  \label{fig:2}
\end{subfigure}\hfil 
\begin{subfigure}{0.2\textwidth}
  \includegraphics[width=\linewidth]{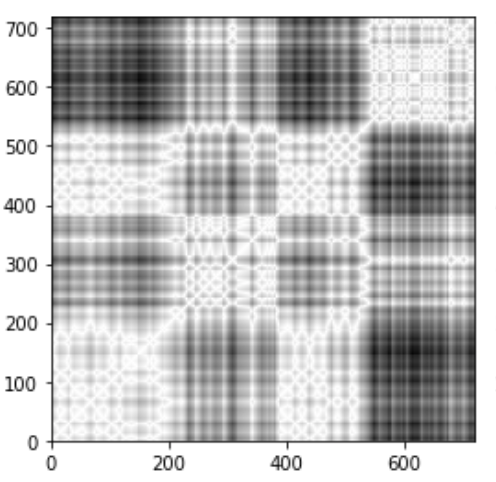}
  \caption*{Actual Temperature}
  \label{fig:3}
\end{subfigure} \hfil
\begin{subfigure}{0.2\textwidth}
  \includegraphics[width=\linewidth]{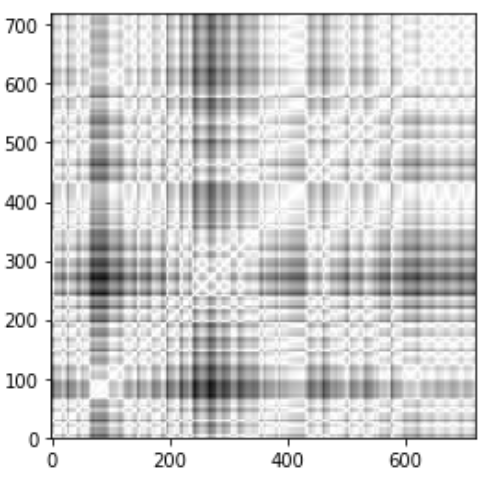}
  \caption*{Air Humidity}
  \label{fig:3}
\end{subfigure}

\medskip
\begin{subfigure}{0.2\textwidth}
  \includegraphics[width=\linewidth]{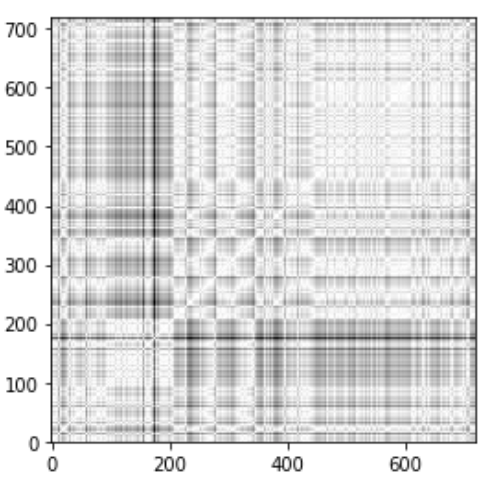}
  \caption*{Relative Humidity}
  \label{fig:4}
\end{subfigure}\hfil 
\begin{subfigure}{0.2\textwidth}
  \includegraphics[width=\linewidth]{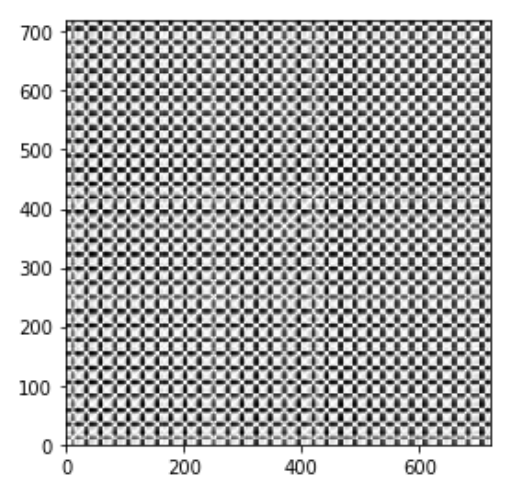}
  \caption*{Brightness}
  \label{fig:5}
\end{subfigure}\hfil 
\begin{subfigure}{0.2\textwidth}
  \includegraphics[width=\linewidth]{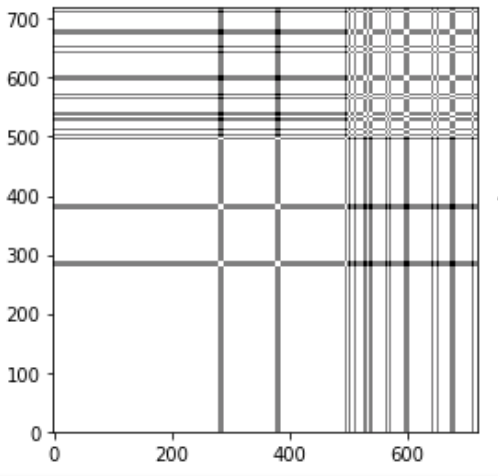}
  \caption*{Motion}
  \label{fig:6}
\end{subfigure} \hfil
\begin{subfigure}{0.2\textwidth}
  \includegraphics[width=\linewidth]{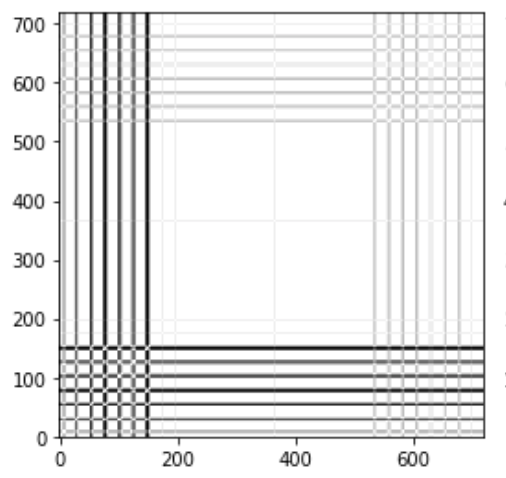}
  \caption*{Intensity}
  \label{fig:6}
\end{subfigure}
\caption{Recurrence plots of time series of different types from the REFIT dataset in grayscale. Notice how the temperature and humidity time series look visually similar.}
\label{recurrenceRefit}
\end{figure*}

\begin{figure*}
\captionsetup[subfigure]{font=scriptsize,labelfont=scriptsize}
    \centering 
\begin{subfigure}{0.2\textwidth}
  \includegraphics[width=\linewidth]{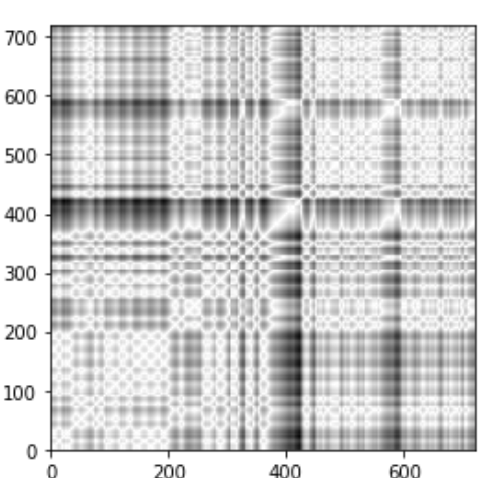}
  \caption*{Air Temp Sensor}
  \label{fig:1}
\end{subfigure}\hfil 
\begin{subfigure}{0.2\textwidth}
  \includegraphics[width=\linewidth]{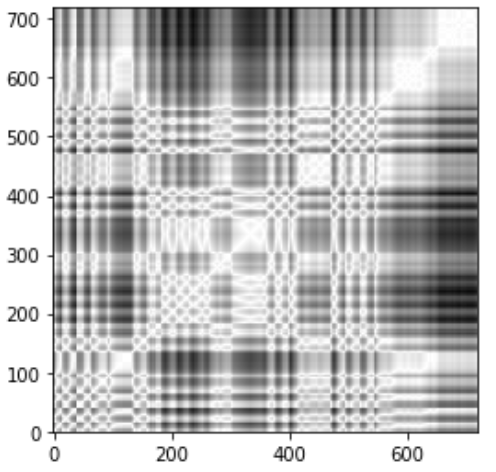}
  \caption*{Temp Sensor}
  \label{fig:2}
\end{subfigure}\hfil 
\begin{subfigure}{0.2\textwidth}
  \includegraphics[width=\linewidth]{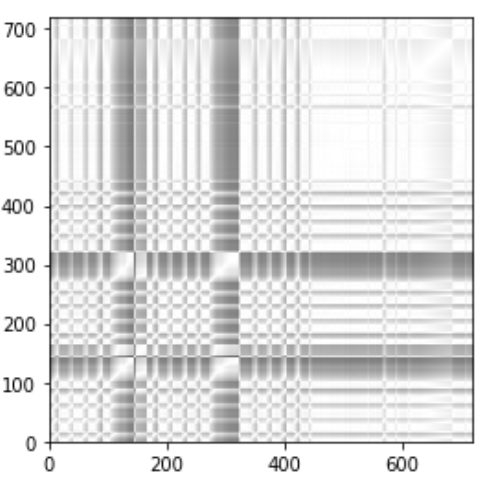}
  \caption*{FCU RTemp Sensor}
  \label{fig:3}
\end{subfigure} \hfil
\begin{subfigure}{0.2\textwidth}
  \includegraphics[width=\linewidth]{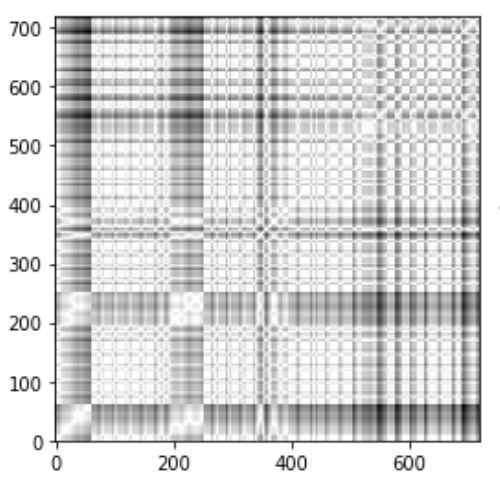}
  \caption*{FCU Temp Sensor}
  \label{fig:3}
\end{subfigure}

\medskip
\begin{subfigure}{0.2\textwidth}
  \includegraphics[width=\linewidth]{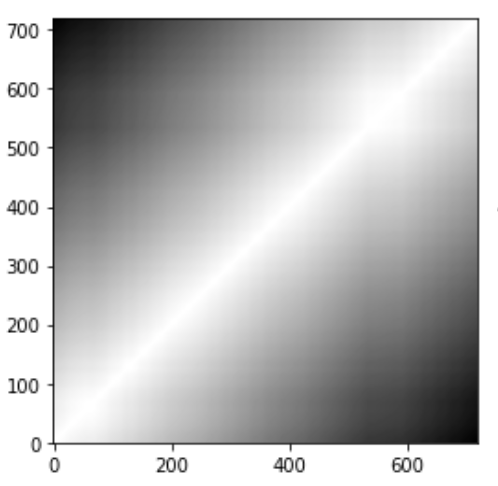}
  \caption*{Energy Meter}
  \label{fig:4}
\end{subfigure}\hfil 
\begin{subfigure}{0.2\textwidth}
  \includegraphics[width=\linewidth]{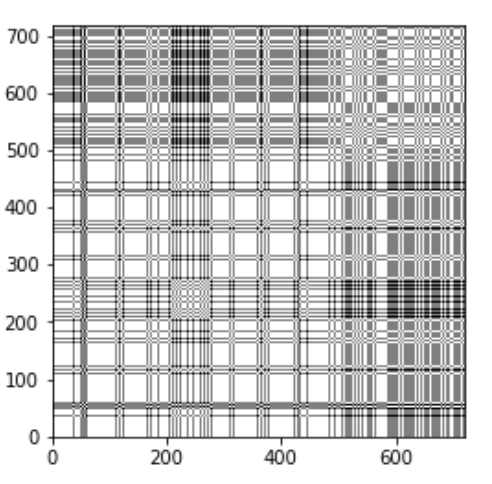}
  \caption*{Load Energy Meter}
  \label{fig:5}
\end{subfigure}\hfil 
\begin{subfigure}{0.2\textwidth}
  \includegraphics[width=\linewidth]{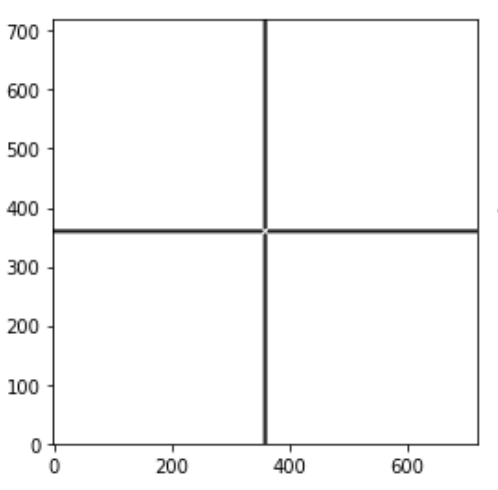}
  \caption*{FCU Fan Speed}
  \label{fig:6}
\end{subfigure} \hfil
\begin{subfigure}{0.2\textwidth}
  \includegraphics[width=\linewidth]{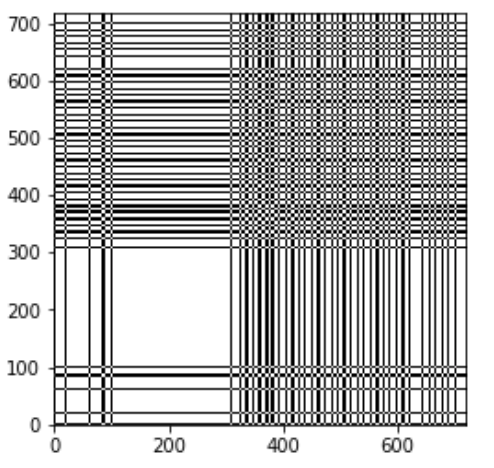}
  \caption*{FCU Flt Sen}
  \label{fig:6}
\end{subfigure}
\caption{Recurrence plots of time series of different types from the Industrial dataset in gray scale. Notice how the \textit{temp} types are distinctively different from the other types but similar to the \textit{Temperature} types in figure \ref{recurrenceRefit}.}
\label{recurrenceDUB}
\end{figure*}

As described in Section~\ref{IETS}, we generated features for the semantic labels in the datasets using the recurrence function. We present the recurrence plots of some of these label in Figures.~\ref{recurrenceRefit} and \ref{recurrenceDUB}. 
Recall that we select only 720 observations from the time series files for all experiments involving IETS. 
Notice how in Fig.~\ref{recurrenceRefit}, there is a clear textural similarity between time series of type \textit{temperature} and \textit{humidity}. Again, we see this similarity in Fig.~\ref{recurrenceDUB} where the time series of type \textit{temp}, exhibit the same textural similarity observed in Fig.~\ref{recurrenceRefit}. The \textit{temperature} time series from both datasets look very similar but distinctly different from the other types. 

We use these encoded representations as features for the supervised learning algorithms. The output of the recurrence function will be a matrix of size $720 \times 720$. This presents a huge computational cost to process and we mitigate this by downsampling. Thus, we proceeded to downsample the image matrix by aggregation to four reduced dimensions: 8, 16, 32 and 48. Then, we transformed these matrices into vectors by flattening before using them as predictors in the algorithms. The same learning methodology described in Section~\ref{supervisedLearning} is adopted here. We present the evaluations of the models trained using IETS in Tables~\ref{macroFscoreBothMethods} and~\ref{methodComparisonBalanedDatasetWithStdev}.

\subsection{Comparison of Methods}
We evaluate the models trained using the descriptive statistical features, IETS features and cosine distances. In all cases, we balanced the datasets by downsampling to the lowest occuring class size. The results presented here are the averages calculated after five runs of each experiments. We present our evaluations in Tables~\ref{macroFscoreBothMethods} and~\ref{methodComparisonBalanedDatasetWithStdev}. 

In Table~\ref{macroFscoreBothMethods}, we present the \textit{macro-averaged} F-score using the descriptive statistical features (DF) and IETS. Best results of both methods for each algorithm is highlighted in bold for emphasis. We can see that for the REFIT dataset, using the descriptive features leads to better results than using IETS. On the other hand, for the industrial dataset, we see that IETS leads to an overall improvement in model performance. Given the semantic closeness of the labels in the industrial dataset, we could deduce that features generated using the IETS encoding are better for discriminating semantically close labels.

In Table~\ref{methodComparisonBalanedDatasetWithStdev}, we present the best average results for each case and the standard deviations alongside. We compare the results for models trained using the descriptive statistical features (DF), IETS features and cosine distances (CD). For the IETS features, we show the results using the reduced dimensions considered: 8, 16, 32 and 48. Furthermore, we show how the models fare across varying training sizes: 20\%, 40\%, 70\% and 80\%. In each case, the test size is a complement of the training size. This experimental methodology is designed to highlight the generalization power of the models. The metric used here is accuracy which is equivalent to the \textit{micro-averaged} F-score for a balanced dataset. Values highlighted in bold are the best results for that case with statistical significance. 
\begin{table}
\centering
  \caption{Table of macro F-score values from using both methods across different algorithms}
  \label{macroFscoreBothMethods}
  \addtolength\tabcolsep{5pt}
  \resizebox{0.7\textwidth}{!}{
  \begin{tabular}{lcccc}
    \toprule
    \multicolumn{1}{c}{\multirow{2}{*}{}}&
    \multicolumn{2}{c}{\multirow{1}{*}{\textbf{REFIT}}}&
     \multicolumn{2}{c}{\multirow{1}{*}{\textbf{Industrial dataset}}}\\\cmidrule(lr{1em}){2-3}\cmidrule(lr{1em}){4-5}
    
    \textbf{Algorithm} & \textbf{DF} & \textbf{IETS} &\textbf{DF} &\textbf{IETS}\\
    \midrule
 \textbf{Logistic Regression} &\textbf{74.5}& 57.6 & 63.0 & \textbf{64.0}  \\
   
\textbf{Random Forests} & \textbf{80.9} & 56.8 & 71.9 & \textbf{82.7} \\

\textbf{Decision Trees} & \textbf{77.4} & 51.2 & \textbf{66.9} & 66.2 \\

\textbf{K-nearest Neighbours} & \textbf{78.5} & 61.8 & 69.9 & \textbf{73.9} \\

\textbf{AdaBoost} & \textbf{79.5} & 66.1 & 77.3 & \textbf{82.0} \\

\textbf{Support Vector Machines} & \textbf{77.3} & 65.1 & 55.1 & \textbf{64.7} \\

    \bottomrule
  \end{tabular}
  }
\end{table}
We see in Table~\ref{methodComparisonBalanedDatasetWithStdev} that the IETS is the clear winner in many cases.

Furthermore, we also see that in cases where it does not win, it is almost as good as the best performing methods for that case or the next best method. For example, in the industrial dataset with just a training size of 20\%, using the features derived from IETS give the best accuracies of 72\% and 68\%, respectively. The significance is further emphasized when we take into consideration the fact that the IETS uses just one month's worth of observations as compared to the others that use the entire dataset. It can also be seen that IETS seems to perform better for the industrial dataset. This may be related to the fact that many of the semantic types are closely related but measured by different devices (e.g. FCU temp vs AirTemp). While the descriptive features may have been good enough for the REFIT dataset, the closeness in similarity exhibited in the industrial dataset can be best discriminated using IETS.

\begin{table}
\centering
\caption{Comparing supervised methods. Metric used here is accuracy (equivalent to the \textit{micro} F-score). Results presented here are the averages taken over five runs, standard deviations are shown alongside. For each method, the results of the best performing algorithm is presented here. The best result for each train size is in \textbf{bold*}. Second best results are \underline{underlined}. Best results represent statistically significant superior values relative to the second best at $\alpha = 0.05$ level of significance using the \textbf{t-test}.}
\begin{subtable}{0.8\textwidth}
\centering
    \label{methodComparisonBalanedDatasetWithStdev}
      \resizebox{0.8\textwidth}{!}{
    \begin{tabular}{lcccc}
    \toprule
    
    \multicolumn{1}{l}{\multirow{1}{*}{}}&
    \multicolumn{4}{c}{\multirow{1}{*}{Train Size (\%)}}\\
    \midrule
    Methods & 20 & 40 & 70 & 80 \\
    \midrule
    with DF & \textbf{66.0*}  $\pm$ 0.02 & \textbf{71.0*}  $\pm$ 0.00 & \textbf{72.0*}  $\pm$ 0.00 & 65.0  $\pm$ 0.02\\ 

    with CD & 54.0  $\pm$ 0.02 & \underline{67.0}  $\pm$ 0.00 & 62.0  $\pm$ 0.00 & \textbf{70.0*}  $\pm$ 0.00\\
    
    
    IETSx8 & 54.0  $\pm$ 0.01 & 61.0  $\pm$ 0.02 & 66.0  $\pm$ 0.00 & 65.0  $\pm$ 0.02 \\

    IETSx16  & 55.0 $\pm$ 0.01 & 65.0  $\pm$ 0.01 & \textbf{72.0*}  $\pm$ 0.00 & 66.0  $\pm$ 0.01\\ 

    IETSx32 & 56.0  $\pm$ 0.02 & 66.0  $\pm$ 0.01 & 69.0  $\pm$ 0.01 & 68.0  $\pm$ 0.01 \\

    IETSx48 & \underline{58.0}  $\pm$ 0.01 & \underline{67.0}  $\pm$ 0.01 & \underline{70.0}  $\pm$ 0.00 & \underline{69.0}  $\pm$ 0.02 \\
    
    \bottomrule
    \end{tabular}
      }
      \caption{REFIT Dataset}\label{REFITdatasetFscore}
\end{subtable}

\bigskip
\begin{subtable}{0.8\textwidth}
    \centering
      \resizebox{0.8\textwidth}{!}{
    \begin{tabular}{lcccc}
    \toprule
    
    \multicolumn{1}{l}{\multirow{1}{*}{}}&
    \multicolumn{4}{c}{\multirow{1}{*}{Train Size (\%)}}\\
    \midrule
    Methods & 20 & 40 & 70 & 80 \\
    \midrule
    with DF & 60.0  $\pm$ 0.03 & 72.0  $\pm$ 0.01 & \textbf{83.0*}  $\pm$ 0.02 & \textbf{77.0*}  $\pm$ 0.04\\

    with CD & 45.0  $\pm$ 0.02 & 52.0  $\pm$ 0.02 & 57.0  $\pm$ 0.00 & 53.0  $\pm$ 0.00\\

    IETSx8 & 54.0  $\pm$ 0.04 & 67.0  $\pm$ 0.00 & 73.0  $\pm$ 0.00& 73.0  $\pm$ 0.00 \\

    IETSx16  & \underline{68.0}  $\pm$ 0.00 & 71.0  $\pm$ 0.02 & \underline{74.0}  $\pm$ 0.00 & \underline{75.0}  $\pm$ 0.00\\

    IETSx32 & \underline{68.0}  $\pm$ 0.00 & \underline{73.0}  $\pm$ 0.02 & \underline{74.0}  $\pm$ 0.00 & 74.0  $\pm$ 0.03 \\

    IETSx48 & \textbf{72.0*}  $\pm$ 0.00 & \textbf{76.0*}  $\pm$ 0.00 & 73.0  $\pm$ 0.02 & \underline{75.0}  $\pm$ 0.04 \\
    \bottomrule
    \end{tabular}
      }
      \caption{Industrial Dataset}\label{industrialDatasetFscore}
\end{subtable}
\label{methodComparisonBalanedDatasetWithStdev}
\end{table}

\section{Conclusions}
\label{conclusions}

In this paper, we conducted a study that investigates the feasibility and performance of different supervised learning methods for inferring the semantics of IoT devices in \textit{smart buildings}. Particularly, we highlight the potential of Image Encoded Time Series (IETS) as an alternative to descriptive statistics features. We show that using just a fraction of the data in a time series file, this encoding can produce features that compete with traditional statistical feature-based methods. This is similar to data-driven paradigms such as budgeted learning~\cite{iddianozie2020exploring,lizotte2012budgeted}, \textit{zero-} and \textit{one-} shot learning~\cite{socher2013zero,norouzi2013zero,fei2006one} that seek to improve learning on small datasets. 

The approaches described in this study will be beneficial to machine learning research endeavours, particularly for IoT in \textit{smart buildings} where the lack of good quality data can sometimes be an impediment~\cite{shi2018survey}. A key takeaway from the evaluations is that a machine learning panacea for addressing the problem of semantic heterogeneity does not exist yet. We see this in the fact that our approaches perform differently with the datasets. For example, the IETS method outperforms the traditional feature-based methods for the industrial dataset (see Table.~\ref{macroFscoreBothMethods}). Thus, these methods could outperform each other in different scenarios. 







We recognize that in this study, we did not incorporate the seasonality in the experimental process. This is a limitation of this study as it is likely that this could affect the results produced from the encoding with regards to different semantic labels. For future work, we will investigate the effect of time series seasonality on the inference methods of IETS. Likewise, we will research the relation between the size of time series points and the performance of the inference models in an attempt to determine a globally optimal size for different semantic types. In addition, we will extend our methods to paradigms like transfer learning as it will be interesting to see how they scale to different tasks and domains~\cite{iddianozie2019transfer,pan2009survey}.

\newpage

\bibliographystyle{unsrt} 

\bibliography{references.bib}

\begin{thebibliography}{10}

\bibitem{UN2018}
United Nations~Department of~Economic and Social Affairs.
\newblock 68\% of the world population projected to live in urban areas by
  2050, says {UN}, 2020.

\bibitem{akande2019lisbon}
Adeoluwa Akande, Pedro Cabral, Paulo Gomes, and Sven Casteleyn.
\newblock The lisbon ranking for smart sustainable cities in europe.
\newblock {\em Sustainable Cities and Society}, 44:475--487, 2019.

\bibitem{leal2018reinvigorating}
Walter Leal~Filho, Ulisses Azeiteiro, F{\'a}tima Alves, Paul Pace, Mark Mifsud,
  Luciana Brandli, Sandra~S Caeiro, and Antje Disterheft.
\newblock Reinvigorating the sustainable development research agenda: the role
  of the sustainable development goals (sdg).
\newblock {\em International Journal of Sustainable Development \& World
  Ecology}, 25(2):131--142, 2018.

\bibitem{lu2015policy}
Yonglong Lu, Nebojsa Nakicenovic, Martin Visbeck, and Anne-Sophie Stevance.
\newblock Policy: five priorities for the un sustainable development goals.
\newblock {\em Nature}, 520(7548):432--433, 2015.

\bibitem{campisi2018evaluation}
Domenico Campisi, Simone Gitto, and Donato Morea.
\newblock An evaluation of energy and economic efficiency in residential
  buildings sector: A multi-criteria analisys on an italian case study.
\newblock {\em International Journal of Energy Economics and Policy}, 8(3):185,
  2018.

\bibitem{silva2018towards}
Bhagya~Nathali Silva, Murad Khan, and Kijun Han.
\newblock Towards sustainable smart cities: A review of trends, architectures,
  components, and open challenges in smart cities.
\newblock {\em Sustainable Cities and Society}, 38:697--713, 2018.

\bibitem{chen2014vision}
Shanzhi Chen, Hui Xu, Dake Liu, Bo~Hu, and Hucheng Wang.
\newblock A vision of iot: Applications, challenges, and opportunities with
  china perspective.
\newblock {\em IEEE Internet of Things journal}, 1(4):349--359, 2014.

\bibitem{ahvenniemi2017differences}
Hannele Ahvenniemi, Aapo Huovila, Isabel Pinto-Sepp{\"a}, and Miimu Airaksinen.
\newblock What are the differences between sustainable and smart cities?
\newblock {\em Cities}, 60:234--245, 2017.

\bibitem{UNECE2015}
United~Nations Economic and Social Council.
\newblock The unece–itu smart sustainable cities indicators, 2020.

\bibitem{kramers2014smart}
Anna Kramers, Mattias H{\"o}jer, Nina L{\"o}vehagen, and Josefin Wangel.
\newblock Smart sustainable cities--exploring ict solutions for reduced energy
  use in cities.
\newblock {\em Environmental modelling \& software}, 56:52--62, 2014.

\bibitem{lilis2017towards}
Georgios Lilis, Gilbert Conus, Nastaran Asadi, and Maher Kayal.
\newblock Towards the next generation of intelligent building: An assessment
  study of current automation and future iot based systems with a proposal for
  transitional design.
\newblock {\em Sustainable cities and society}, 28:473--481, 2017.

\bibitem{jung2012integrating}
Markus Jung, Christian Reinisch, and Wolfgang Kastner.
\newblock Integrating building automation systems and ipv6 in the internet of
  things.
\newblock In {\em 2012 Sixth International Conference on Innovative Mobile and
  Internet Services in Ubiquitous Computing}, pages 683--688. IEEE, 2012.

\bibitem{akkaya2015iot}
Kemal Akkaya, Ismail Guvenc, Ramazan Aygun, Nezih Pala, and Abdullah Kadri.
\newblock Iot-based occupancy monitoring techniques for energy-efficient smart
  buildings.
\newblock In {\em 2015 IEEE Wireless communications and networking conference
  workshops (WCNCW)}, pages 58--63. IEEE, 2015.

\bibitem{brambley2005advanced}
Michael~R Brambley, Philip Haves, Sean~C McDonald, Paul Torcellini, D~Hansen,
  DR~Holmberg, and Kurt~W Roth.
\newblock Advanced sensors and controls for building applications: Market
  assessment and potential r\&d pathways.
\newblock Technical report, EERE Publication and Product Library, Washington,
  DC (United States), 2005.

\bibitem{shaikh2014review}
Pervez~Hameed Shaikh, Nursyarizal Bin~Mohd Nor, Perumal Nallagownden, Irraivan
  Elamvazuthi, and Taib Ibrahim.
\newblock A review on optimized control systems for building energy and comfort
  management of smart sustainable buildings.
\newblock {\em Renewable and Sustainable Energy Reviews}, 34:409--429, 2014.

\bibitem{effects2018yougov}
Velux.
\newblock The effects of modern indoor living on health, wellbeing and
  productivity.
\newblock Technical report, YouGov, 2018.

\bibitem{bhattacharya2015automated}
Arka~A Bhattacharya, Dezhi Hong, David Culler, Jorge Ortiz, Kamin Whitehouse,
  and Eugene Wu.
\newblock Automated metadata construction to support portable building
  applications.
\newblock In {\em Proceedings of the 2nd ACM International Conference on
  Embedded Systems for Energy-Efficient Built Environments}, pages 3--12. ACM,
  2015.

\bibitem{gao2015data}
Jingkun Gao, Joern Ploennigs, and Mario Berges.
\newblock A data-driven meta-data inference framework for building automation
  systems.
\newblock In {\em Proceedings of the 2nd ACM International Conference on
  Embedded Systems for Energy-Efficient Built Environments}, pages 23--32. ACM,
  2015.

\bibitem{mehta2013sustainable}
D~Paul Mehta and Martin Wiesehan.
\newblock Sustainable energy in building systems.
\newblock {\em Procedia Computer Science}, 19:628--635, 2013.

\bibitem{koh2018scrabble}
Jason Koh, Bharathan Balaji, Dhiman Sengupta, Julian McAuley, Rajesh Gupta, and
  Yuvraj Agarwal.
\newblock Scrabble: transferrable semi-automated semantic metadata
  normalization using intermediate representation.
\newblock In {\em Proceedings of the 5th Conference on Systems for Built
  Environments}, pages 11--20. ACM, 2018.

\bibitem{shi2018survey}
Feifei Shi, Qingjuan Li, Tao Zhu, and Huansheng Ning.
\newblock A survey of data semantization in internet of things.
\newblock {\em Sensors}, 18(1):313, 2018.

\bibitem{haystack2020}
Haystack.
\newblock Project haystack, 2020.

\bibitem{schumann2014towards}
Anika Schumann, Joern Ploennigs, and Bernard Gorman.
\newblock Towards automating the deployment of energy saving approaches in
  buildings.
\newblock In {\em Proceedings of the 1st ACM Conference on Embedded Systems for
  Energy-Efficient Buildings}, pages 164--167, 2014.

\bibitem{koc2014comparison}
Merthan Koc, Burcu Akinci, and Mario Berg{\'e}s.
\newblock Comparison of linear correlation and a statistical dependency measure
  for inferring spatial relation of temperature sensors in buildings.
\newblock In {\em Proceedings of the 1st ACM Conference on Embedded Systems for
  Energy-Efficient Buildings}, pages 152--155, 2014.

\bibitem{fulcher2014highly}
Ben~D Fulcher and Nick~S Jones.
\newblock Highly comparative feature-based time-series classification.
\newblock {\em IEEE Transactions on Knowledge and Data Engineering},
  26(12):3026--3037, 2014.

\bibitem{hyndman2015large}
Rob~J Hyndman, Earo Wang, and Nikolay Laptev.
\newblock Large-scale unusual time series detection.
\newblock In {\em 2015 IEEE international conference on data mining workshop
  (ICDMW)}, pages 1616--1619. IEEE, 2015.

\bibitem{kolozali2016effect}
{\c{S}}efki Kolozali, Daniel Puschmann, Maria Bermudez-Edo, and Payam Barnaghi.
\newblock On the effect of adaptive and nonadaptive analysis of time-series
  sensory data.
\newblock {\em IEEE Internet of Things Journal}, 3(6):1084--1098, 2016.

\bibitem{ye2009time}
Lexiang Ye and Eamonn Keogh.
\newblock Time series shapelets: a new primitive for data mining.
\newblock In {\em Proceedings of the 15th ACM SIGKDD international conference
  on Knowledge discovery and data mining}, pages 947--956. ACM, 2009.

\bibitem{hatami2018classification}
Nima Hatami, Yann Gavet, and Johan Debayle.
\newblock Classification of time-series images using deep convolutional neural
  networks.
\newblock In {\em Tenth International Conference on Machine Vision (ICMV
  2017)}, volume 10696, page 106960Y. International Society for Optics and
  Photonics, 2018.

\bibitem{calbimonte2012deriving}
Jean-Paul Calbimonte, Oscar Corcho, Zhixian Yan, H~Jeung, and Karl Aberer.
\newblock Deriving semantic sensor metadata from raw measurements.
\newblock 2012.

\bibitem{gonzalez2018beats}
Aurora Gonzalez-Vidal, Payam Barnaghi, and Antonio~F Skarmeta.
\newblock Beats: Blocks of eigenvalues algorithm for time series segmentation.
\newblock {\em IEEE Transactions on Knowledge and Data Engineering},
  30(11):2051--2064, 2018.

\bibitem{paparrizos2019grail}
John Paparrizos and Michael~J Franklin.
\newblock Grail: efficient time-series representation learning.
\newblock {\em Proceedings of the VLDB Endowment}, 12(11):1762--1777, 2019.

\bibitem{eckmann1995recurrence}
JP~Eckmann, S~Oliffson Kamphorst, D~Ruelle, et~al.
\newblock Recurrence plots of dynamical systems.
\newblock {\em World Scientific Series on Nonlinear Science Series A},
  16:441--446, 1995.

\bibitem{wang2015encoding}
Zhiguang Wang and Tim Oates.
\newblock Encoding time series as images for visual inspection and
  classification using tiled convolutional neural networks.
\newblock In {\em Workshops at the Twenty-Ninth AAAI Conference on Artificial
  Intelligence}, 2015.

\bibitem{Firth2017}
Steven Firth, Tom Kane, Vanda Dimitriou, Tarek Hassan, Farid Fouchal, Michael
  Coleman, and Lynda Webb.
\newblock {REFIT Smart Home dataset}.
\newblock 6 2017.

\bibitem{blu2004linear}
Thierry Blu, Philippe Th{\'e}venaz, and Michael Unser.
\newblock Linear interpolation revitalized.
\newblock {\em IEEE Transactions on Image Processing}, 13(5):710--719, 2004.

\bibitem{breiman2001random}
Leo Breiman.
\newblock Random forests.
\newblock {\em Machine learning}, 45(1):5--32, 2001.

\bibitem{hosmer2013applied}
David~W Hosmer~Jr, Stanley Lemeshow, and Rodney~X Sturdivant.
\newblock {\em Applied logistic regression}, volume 398.
\newblock John Wiley \& Sons, 2013.

\bibitem{quinlan1986induction}
J.~Ross Quinlan.
\newblock Induction of decision trees.
\newblock {\em Machine learning}, 1(1):81--106, 1986.

\bibitem{cover1967nearest}
Thomas Cover and Peter Hart.
\newblock Nearest neighbor pattern classification.
\newblock {\em IEEE transactions on information theory}, 13(1):21--27, 1967.

\bibitem{hastie2009multi}
Trevor Hastie, Saharon Rosset, Ji~Zhu, and Hui Zou.
\newblock Multi-class adaboost.
\newblock {\em Statistics and its Interface}, 2(3):349--360, 2009.

\bibitem{ng2002spectral}
Andrew~Y Ng, Michael~I Jordan, and Yair Weiss.
\newblock On spectral clustering: Analysis and an algorithm.
\newblock In {\em Advances in neural information processing systems}, pages
  849--856, 2002.

\bibitem{park2009simple}
Hae-Sang Park and Chi-Hyuck Jun.
\newblock A simple and fast algorithm for k-medoids clustering.
\newblock {\em Expert systems with applications}, 36(2):3336--3341, 2009.

\bibitem{hartigan1979algorithm}
John~A Hartigan and Manchek~A Wong.
\newblock Algorithm as 136: A k-means clustering algorithm.
\newblock {\em Journal of the Royal Statistical Society. Series C (Applied
  Statistics)}, 28(1):100--108, 1979.

\bibitem{manning2010introduction}
Christopher Manning, Prabhakar Raghavan, and Hinrich Sch{\"u}tze.
\newblock Introduction to information retrieval.
\newblock {\em Natural Language Engineering}, 16(1):100--103, 2010.

\bibitem{maaten2008visualizing}
Laurens van~der Maaten and Geoffrey Hinton.
\newblock Visualizing data using t-sne.
\newblock {\em Journal of machine learning research}, 9(Nov):2579--2605, 2008.

\bibitem{iddianozie2020exploring}
Chidubem Iddianozie, Michela Bertolotto, and Gavin Mcardle.
\newblock Exploring budgeted learning for data-driven semantic inference via
  urban functions.
\newblock {\em IEEE Access}, 8:32258--32269, 2020.

\bibitem{lizotte2012budgeted}
Daniel~J Lizotte, Omid Madani, and Russell Greiner.
\newblock Budgeted learning of naive-bayes classifiers.
\newblock {\em arXiv preprint arXiv:1212.2472}, 2012.

\bibitem{socher2013zero}
Richard Socher, Milind Ganjoo, Christopher~D Manning, and Andrew Ng.
\newblock Zero-shot learning through cross-modal transfer.
\newblock In {\em Advances in neural information processing systems}, pages
  935--943, 2013.

\bibitem{norouzi2013zero}
Mohammad Norouzi, Tomas Mikolov, Samy Bengio, Yoram Singer, Jonathon Shlens,
  Andrea Frome, Greg~S Corrado, and Jeffrey Dean.
\newblock Zero-shot learning by convex combination of semantic embeddings.
\newblock {\em arXiv preprint arXiv:1312.5650}, 2013.

\bibitem{fei2006one}
Li~Fei-Fei, Rob Fergus, and Pietro Perona.
\newblock One-shot learning of object categories.
\newblock {\em IEEE transactions on pattern analysis and machine intelligence},
  28(4):594--611, 2006.

\bibitem{iddianozie2019transfer}
Chidubem Iddianozie and Gavin McArdle.
\newblock A transfer learning paradigm for spatial networks.
\newblock In {\em Proceedings of the 34th ACM/SIGAPP Symposium on Applied
  Computing}, pages 659--666, 2019.

\bibitem{pan2009survey}
Sinno~Jialin Pan and Qiang Yang.
\newblock A survey on transfer learning.
\newblock {\em IEEE Transactions on knowledge and data engineering},
  22(10):1345--1359, 2009.

\end{thebibliography}






\end{document}